\documentclass[12pt,preprint]{aastex}

\newcommand{\etal}{et~al.}

\newcommand{\ks}{$K_{\rm s}$}

\newcommand{\msun}{M$_{\sun}$}

\newcommand{\mum}{$\mu$m}
\def\asec{$^{\prime\prime}~$}
\def\amin{$^{\prime}$~}
\def\deg{$^{\circ}~$}

\begin{document}

\title{The Spitzer c2d Survey of Large, Nearby, Interstellar Clouds: \\
VII.  Ophiuchus Observed with MIPS}

\author{Deborah L.\ Padgett\altaffilmark{1,2},
Luisa M.\ Rebull\altaffilmark{1},
Karl R.\ Stapelfeldt\altaffilmark{3},
Nicholas L.\ Chapman\altaffilmark{4},
Shih-Ping Lai\altaffilmark{4,5,6},
Lee G.\ Mundy\altaffilmark{4},
Neal J.\ Evans II\altaffilmark{7},
Timothy Y.\ Brooke\altaffilmark{1},
Lucas A.\ Cieza\altaffilmark{7},
William J.\ Spiesman\altaffilmark{7},
Alberto Noriega-Crespo\altaffilmark{1},
Caer-Eve McCabe\altaffilmark{1,3}
Lori E.\ Allen\altaffilmark{8},
Geoffrey A.\ Blake\altaffilmark{9},
Paul M.\ Harvey\altaffilmark{7},
Tracy L.\ Huard\altaffilmark{8},
Jes K.\ J{\o}rgensen\altaffilmark{8,10},
David W.\ Koerner\altaffilmark{11},
Philip C.\ Myers\altaffilmark{8},
Annelia I.\ Sargent\altaffilmark{12},
Peter Teuben\altaffilmark{4},
Ewine F.\ van Dishoeck\altaffilmark{13},
Zahed Wahhaj\altaffilmark{11,14},
\& Kaisa E.\ Young\altaffilmark{7,15}}

\altaffiltext{1}{Spitzer Science Center, MC 220-6,
California Institute of Technology, Pasadena, CA 91125 USA}
\altaffiltext{2}{email: dlp@ipac.caltech.edu}
\altaffiltext{3}{Jet Propulsion Laboratory, MS 183-900,
4800 Oak Grove Drive, Pasadena, CA 91109}
\altaffiltext{4}{Department of Astronomy, University of 
Maryland, College Park, MD 20742}
\altaffiltext{5}{Institute of Astronomy and Department of 
Physics, National Tsing Hua University, Hsinchu 30043, Taiwan}
\altaffiltext{6}{Academia Sinica Institute of Astronomy and 
Astrophysics, P.O. Box 23-141, Taipei 106, Taiwan}
\altaffiltext{7}{Department of Astronomy, University of 
Texas at Austin, 1 University Station C1400, Austin, TX 78712}
\altaffiltext{8}{Harvard-Smithsonian Center for 
Astrophysics, 60 Garden Street, MS42, Cambridge, MA 02138}
\altaffiltext{9}{Division of Geological and Planetary 
Sciences, MS 150-21, California Institute of Technology, 
Pasadena, CA 91125}
\altaffiltext{10}{Current address: Argelander-Institut für Astronomie
Auf dem Hügel 71 D-53121 Bonn Germany.}
\altaffiltext{11}{Department of Physics and Astronomy,
Northern Arizona University, NAU Box 6010, Flagstaff, AZ 86011-6010}
\altaffiltext{12}{Division of Physics, Mathematics, and
Astronomy, MS 105-24, California Institute of Technology, 
Pasadena, CA 91125}
\altaffiltext{13}{Leiden Observatory, PO Box 9513, NL 2300 
RA Leiden, The Netherlands}
\altaffiltext{14}{Current address: Institute for Astronomy, 
University of Hawaii, 2680 Woodlawn Drive, Honolulu HI 96822-1897}
\altaffiltext{15}{Current address: Department of Physical Sciences, 
Nicholls State University, Thibodaux, Louisiana 70301}

\begin{abstract}

We present maps of 14.4 deg$^2$ of the Ophiuchus dark clouds
observed by the {\it Spitzer} Space Telescope Multiband Imaging
Photometer for {\it Spitzer} (MIPS).  These high quality maps
depict both numerous point sources as well as extended dust
emission within the star-forming and non-star-forming portions
of these clouds.  Using PSF-fitting photometry, we detect 5779 sources
at 24 \mum\ and 81 sources at 70 \mum\ at the 10$\sigma$ level of
significance.  Three hundred twenty-three candidate young stellar
objects (YSOs) were identified according to their positions on the
MIPS/2MASS K versus K$-$[24] color-magnitude diagrams as compared
to 24 \mum\ detections in the SWIRE extragalactic survey. 
We find that more than half of the YSO candidates, and almost all
the ones with protostellar Class I spectral energy distributions,
are confined to the known cluster and aggregates.


\end{abstract}
    

\section{Introduction}

The Ophiuchus molecular cloud complex is one of the most well-known
regions of nearby active star formation. For the Lynds
1688 cloud alone, {\it Simbad} lists 770 references as of summer
2006. The clouds ($\alpha$ = 16$^h$28$^m$06$^s$, $\delta$ =
$-$24$\arcdeg$32.5$\arcmin$) are considered to be at a distance
of 125$\pm$25 pc (de Geus et al.\ 1989), although  more distant
estimates are seen in the literature (145 pc, de Zeeuw et al.\
1999; 160 pc, Chini 1981). The stellar population of the
Ophiuchus clouds has been the subject of many imaging surveys 
(Grasdalen, Strom, \& Strom 1973; Vrba et al.\ 1975; Wilking \& Lada 
1983; Lada \& Wilking 1984; Rieke et al.\ 1989; Strom et al.\ 1995; 
Barsony et al.\ 1989, 1997, 2005; Greene et al.\ 1992; Comeron et al.\ 1993; 
Bontemps et al.\ 1998; Allen et al.\ 2002;  Duch\^ene et al.\ 2004; etc.) 
and spectroscopic surveys (Bouvier \& Appenzeller 1992; Greene et al.\ 
1995; Mart\'in et al.\ 1998; Wilking et al.\ 1999; Luhman et al.\ 1999; 
Cushing et al.\ 2000; Wilking et al.\ 2005; Natta et al.\ 2006; etc.). 
Many thousands of sources have been cataloged at near-IR
wavelengths; however, the majority of these sources are
extincted background stars (Barsony et al.\ 1997). 
   
   L1688 contains the most famous star-forming cluster in this region
with of order 100 stars $<$1 Myr old, including several extremely
young Class 0 - I sources (Bontemps et al.\ 2001; Duch\^ene et al.\
2004;  Haisch et al.\ 2002; and many older studies). The
vicinity of this  cluster is covered in bright and dark
nebulosity, and Herbig-Haro flows criss-cross the entire region
(Gomez, Whitney, \& Wood 1998; Phelps et al.\ 2006). Most of 
the known young stellar objects are found within an embedded
cluster only a few parsecs in extent, making Ophiuchus one of
the higher density environments among the nearby low mass star
formation regions (Grasdalen, Strom, \& Strom 1973; Wilking \&
Lada 1983; Wilking \& Lada 1989, etc.).  It is also a region
forming intermediate mass stars, bridging the gap between Taurus
(a region of low-mass, isolated star formation) and Orion (a dense 
cluster with massive O stars).  In Ophiuchus, the principal region 
of low mass star formation is in the vicinity of several intermediate 
to high mass stars of the Sco OB1 association (de Zeeuw 1999), and a 
number of A and B stars appear to be embedded within the cloud 
(Klose 1986; Devito \& Haywood 1998).  

   The molecular material 
in the Lynds 1688 and 1689 clouds extends far beyond the boundaries of 
the extant surveys for YSOs. To the southeast of L1688 is the L1689
cloud, which extends to the east in the L1712 and L1729
streamers. Only a few isolated Young Stellar Objects (YSOs) are known 
in this fairly massive cloud, leading Nutter, Ward-Thompson, \& Andr\'e 
(2006) to dub it ``the dog that didn't bark.'' Their submm continuum
mapping survey indicates that most of the mass of L1689 resides
in distributed cloud material rather than dense cores, unlike
L1688, and suggests that the proximity of L1688 to $\sigma$ Sco 
explains its much greater star formation efficiency. Filamentary
clouds (L1709, L1740, and L1744+L1755+L1765 = ``Oph North'')
also extend to the northeast of L1688. To date, these clouds are
lacking in high resolution  mid-IR surveys to identify and
characterize their YSO content. Only the IRAS and 2MASS surveys
exist for these regions. The ISOCAM survey of Oph covered only
0.7 square degrees (Bontemps et al.\ 1998). Those authors claim
that most of the YSOs outside of this central region are
weak-line and post T Tauri stars and, indeed, this is generally
true based on  only the IRAS and ROSAT studies of the region.  The
current study using {\it Spitzer} is capable of detecting low luminosity 
Class 0 - III infrared sources throughout a much larger surveyed area.

The Ophiuchus molecular cloud complex is one of five
star-forming regions observed with the {\it Spitzer}
Space Telescope ({\it Spitzer}) as a part of the ``Molecular
Cores to Planet- forming Disks'' (c2d) Legacy project (Evans et
al.\ 2003). The goal of this survey is to characterize the circumstellar
material of young stars and substellar objects in these clouds.
This paper is one in a series which present the c2d large cloud mapping
results made with {\it Spitzer's} Infrared Camera (IRAC) and Multi-band 
Imaging Photometer for {\it Spitzer} (MIPS). Others in the series report
results for Chamaeleon II (MIPS: Young et al.\ 2005; IRAC: Porras
et al.\ 2007), Lupus (MIPS: Chapman et al.\ 2007), Perseus (MIPS: 
Rebull et al.\ 2007; IRAC: J{\o}rgensen et al.\ 2006), and Serpens 
(IRAC: Harvey et al.\ 2006; MIPS: Harvey et al.\ 2007).  These papers 
endeavor to present the results for each cloud and instrument in a 
standard format.  Although the current work principally discusses the  
Ophiuchus MIPS maps, data from IRAC are also integrated into sections 
regarding source classification and individual objects. The IRAC results 
for Ophiuchus will be presented in Allen et al., in preparation.

The c2d MIPS survey of Ophiuchus attempts to determine the
importance of a more distributed mode of star formation in the
less-studied areas outside the Oph core by covering more than
13.8 square degrees in the region where $A_V \geq$ 5 at
sensitivity levels considerably better than that of ISOCAM at
mid-IR wavelengths and many orders of magnitude better than IRAS
and ISOPHOT at far-IR wavelengths.  The three MIPS filters at
24, 70, and 160 \mum\ are ideally suited for linking the
circumstellar environment of individual stars to the
interstellar environment of the cloud. The MIPS 24 \mum\ band
is especially useful for identifying stars with infrared excess
since stellar photospheric emission has
decreased by two orders of magnitude from the blackbody peak,
increasing the contrast with peak emission from cool dust. 
Thus, stars with warm young disks are generally extremely bright
at 24 \mum\ relative to stars without disk emission. MIPS 70
\mum\ emission is also seen in the brighter envelopes and disks and in
interstellar regions where dust is heated by young stars. MIPS
160 \mum\ traces cool dust in both circumstellar and
interstellar regions, similar to the IRAS 100 \mum\
band. The combination of MIPS data with surveys at
other wavelengths enables detailed studies of YSO spectral energy
distributions (SEDs) and the cloud environment.  The most recent 
millimeter continuum maps of this region have been presented by 
Young et al.\ (2006), who used CSO/Bolocam to map the region observed 
by {\it Spitzer}. High quality millimeter molecular line maps of 
Ophiuchus also covering the {\it Spitzer} survey are available from 
the COMPLETE team (Ridge et al.\ 2006).

In this paper, we first present the {\it Spitzer} MIPS
observations of Ophiuchus (\S\ref{sec:obs}). Following the
observations section, we discuss results, including
multi-wavelength images (\S\ref{sec:images}), MIPS source counts
(\S\ref{sec:sourcecounts}),  bandmerging with IRAC and 2MASS
(\S\ref{sec:bm}), and a variety of color-color and
color-magnitude diagrams (\S\ref{sec:cmdccd}).  Then we
investigate recovery of IRAS objects (\S\ref{sec:iras}) 
and variability at 24 \mum\ on timescales of 3-8 hours in
Section~\ref{sec:variability}.  We have highlighted the spectral
energy distributions (SEDs) and imaging of several individual
objects in Section~\ref{sec:indobj}. We then discuss 
the distribution of infrared excess sources in Ophiuchus and
comment on the degree of clustering found throughout the
region (\S\ref{sec:cluster}).  Finally, we summarize
our main points in \S\ref{sec:concl}.

\section{Observations and Data Reduction}
\label{sec:obs}
\subsection{MIPS}

The MIPS observations of Ophiuchus were conducted on 2004 March
18-19 and covered 13.8 square degrees.  Two off-cloud fields totalling 
0.6 square degrees were observed with the intention of characterizing 
the background starcounts.  The center of this large map is roughly at 
$\alpha, \delta$ (J2000) = 16$^h$33$^m$18.0$^s$ $-$24$\arcdeg$15$\arcmin$, 
or $l,b$=354$\arcdeg$, $+$16.0$\arcdeg$, or ecliptic coordinates (J2000) 
251$\arcdeg$, $-$2.5$\arcdeg$.  The area mapped with MIPS completely covers 
the c2d IRAC map regions in Ophiuchus, which included all the cloud
area above the $A_V$=5 contour. The combination of the cloud being 
extended perpendicular to the MIPS scan direction, and the fixed scan 
length options available for use with MIPS, dictated that the area covered 
by the Ophiuchus MIPS maps is more than twice the area mapped with IRAC.

These observations were part of Spitzer program id 177; the
AORKEYs appear in Table 1. Fast scan maps were obtained at two separate 
epochs separated by a few hours.  At each epoch, the
spacing between adjacent scan legs was 240$\arcsec$.  The
second epoch observation was offset 125$\arcsec$ in the
cross-scan direction from the first, to fill in the 70 \mum\ sky
coverage which was incomplete at each individual epoch. 
Furthermore, the second epoch scan was also offset 80$\arcsec$
from the first in the scan direction, to maximize 160 $\mu$m
map coverage.  These mapping parameters resulted in every part
of the map being imaged at two epochs at 24 \mum\ and only one
epoch at 70 and 160 \mum, with total integration times of 30
sec, 15 sec, and 3 sec at each point in the map (respectively). 
Even with this strategy, the 160 \mum\ maps have some gaps,
which makes point source photometry especially uncertain at this
wavelength. In addition, the 160 \mum\ array saturated in the
region at the center of the L1688 cloud.  

\begin{figure}
\epsscale{0.75}
\plotone{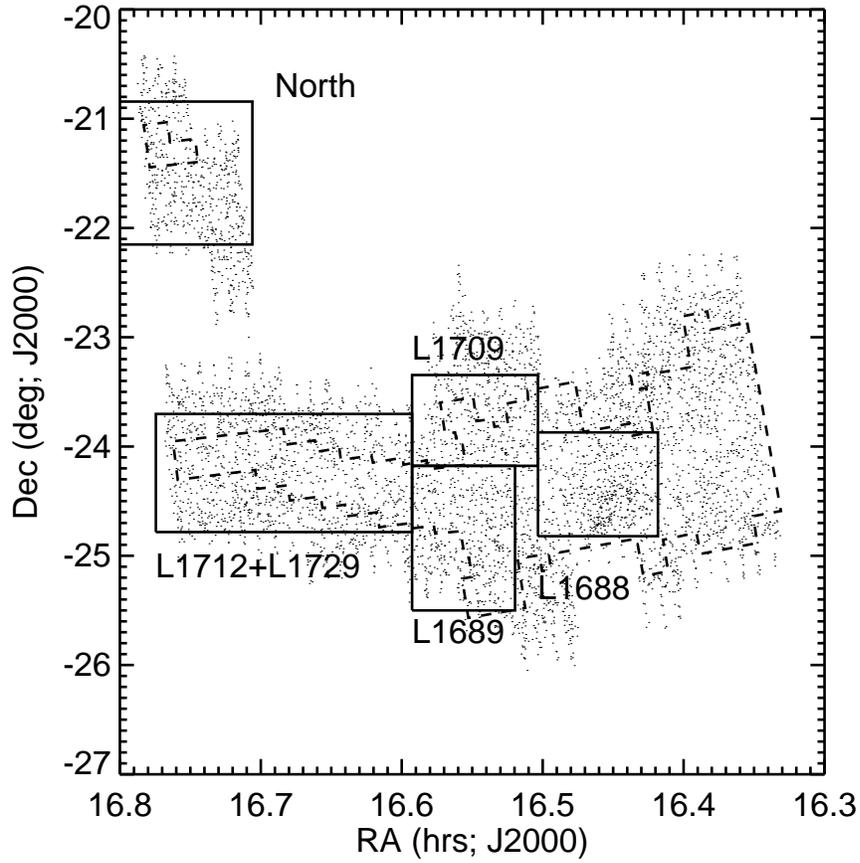}
\caption{Location of MIPS coverage (points are MIPS-24
detections), with the region of IRAC coverage (dashed line) indicated.  
The smaller squares (solid line) indicates the
regions defined to be L1688, L1689, and Ophiuchus North. }
\label{fig:where}
\end{figure}

Figure~\ref{fig:where} shows the region of 3-band coverage with
MIPS, and the 4-band coverage with IRAC.  In addition to the c2d
data, smaller MIPS maps of Ophiuchus were obtained by the {\it Spitzer} 
Guaranteed Time Observers.  To maintain uniform sensitivity in our
maps, these data were not incorporated into the current study.

The observations were conducted at two separate epochs separated
by about 3 to 8 hrs to permit asteroid removal in this low
ecliptic latitude ($\sim0\arcdeg$) field. Study of the asteroid
population in Taurus at a similar ecliptic latitude (0 - 7 deg) 
indicates there are roughtly 250 asteroids per square degree 
with 24 \mum\ flux densities $\ge$ 1.6 mJy 
(Stapelfeldt et al.\ 2006).  By extracting sources separately 
from the two epochs and comparing the source lists, transient sources 
were eliminated from our MIPS catalog. This observation technique also 
enables a study of 24 \mum\ source variability on a timescale of several 
hours; see Section~\ref{sec:variability}.  

The MIPS observations of Ophiuchus included two small off-cloud regions 
with total area of 0.66 square degrees, in addition to the 13.8 square
degrees of the main map. These were intended to provide statistics on 
the stellar and extragalactic background.  However, these regions were 
found to be too small and sparse to yield reliable statistics. In addition, 
one of the fields inadvertantly included the large globular cluster M4, 
which severly biases its source counts. Thus, we do not utilize these
offcloud regions; instead, we compare our statistics to the 6.15 sq.\ 
deg.\ the ELAIS N1 field mapped by the Spitzer Wide-area InfraRed 
Extragalactic (``SWIRE'') survey (Shupe et al.\ 2006), which was re-reduced 
and analyzed using the c2d pipeline. In the low ecliptic latitude Ophiuchus 
region, MIPS source counts are dominated by asteroids and the extragalactic 
background rather than stellar photospheres, and thus a comparison
to the SWIRE field is invaluable despite its considerably different 
Galactic latitude.  However, we do include the off-cloud fields  
(OC1,2) in the table of statistics (see Section~\ref{sec:sourcecounts}).

We started with the SSC-pipeline produced basic calibrated data
(BCDs), version S11.4. Each MIPS channel was then processed
differently and therefore is discussed separately below.
Additional detail on the data reduction process can be found
in Young et al.\ (2005), Chapman et al.\ (2007), and Rebull
et al.\ (2007).

Standard c2d pipeline processing was used for MIPS-24 (Young
\etal\ 2005, Evans 2005). After some corrections for
non-uniform readout variations (jailbars) were applied to
individual frames (``basic calibrated data'' or BCDs), these
data were transformed into single epoch mosaics using the SSC
MOPEX software (following Chapman et al.\ 2007). Source lists were
extracted  using ``c2dphot'' (Harvey et al.\ 2006) to measure
flux densities using point spread function (PSF) fitting.  The final working 
2005 catalog as assembled consists of all detections at MIPS-24 
with high quality c2d catalog detection flags (Evans et al.\ 2005)
and detections at both epochs.  While resulting in a
shallower survey than  would be possible using other
combinations of flags, this ensures that no asteroids are
included in the final catalog. The zero point used to convert
these flux densities to magnitudes was 7.14 Jy, based on the
extrapolation from the Vega spectrum as published in the MIPS
Data Handbook. There were 5779 total point sources detected at
24 \mum\ meeting our criteria, ranging from 0.672 to 3910 mJy.
Note that the spatial resolution of MIPS-24 is $\sim6\arcsec$,
with a 2.55$\arcsec$ pixel size.

For the MIPS-70 data, we started with the automated
pipeline-produced BCDs, both the filtered and unfiltered
products.  Then individual BCDs were mosaicked into one
filtered and one unfiltered mosaic using the SSC software MOPEX
(Makovoz et al. 2006).
We performed source detection and extraction using MOPEX/APEX on the
entire mosaic.  The source list was cleaned for instrumental
artifacts via manual inspection of the 70 \mum\ image and
comparison to the 24 \mum\ image; e.g., if there was some
question as to whether a faint object seen at 70 \mum\ was real
or an instrumental artifact, and comparison to the 24 \mum\
image revealed a 24 \mum\ source, then the 70 \mum\ object was
judged to be a real source.  About 80\% of the 70 \mum\ objects 
had identifiable counterparts at 24 \mum. The zero point used to
convert the flux densities to magnitudes was 0.775 Jy, based on the
extrapolation from the Vega spectrum as published in the MIPS
Data Handbook. There were 81 total point sources detected at 70
\mum. 
There were 19 objects brighter than 4 Jy, for which we employed
aperture photometry with a 32$\arcsec$ aperture and a 17\% aperture 
correction.  Note that the spatial resolution of MIPS-70 is 
$\sim18\arcsec$.

The 160 \mum\ MIPS band presented challenges for point source
extraction due to scattered gaps in the map coverage.  A small
amount of smoothing was applied to the data in order to fill
gaps. Then APEX was used to find sources thoughout the image.
Because of the large saturated region in L1688 and the highly
structured nature of the bright diffuse emission, few isolated
point sources were identified in our study. The zero point used
to convert the flux densities to magnitudes was 0.159 Jy, based on the
extrapolation from the Vega spectrum as published in the MIPS
Data Handbook. There were 8 total point sources detected  at 160
\mum, ranging from 876 mJy to 19 Jy.  About 60\% of  the 160
\mum\ objects had identifiable counterparts at 24 and 70 \mum. 
The extreme brightness and saturation of the 160 \mum\ diffuse
emission made detection of 160 \mum\ sources impossible in the
most densely populated regions of the cluster. Finally, note
that the spatial resolution of MIPS-160 is $\sim36\arcsec$.

Although this paper is primarily about point sources, there is
extensive extended emission in all three MIPS channels
throughout the MIPS maps. The diffuse emission ranges in
brightness from 50 -- 570 MJy/sr at 24 \mum\ where the low value
is set by the  overall zodiacal emission in this low ecliptic
latitude region. The 70 \mum\ extended emission ranges in surface 
brightness from $\sim$30 -- 5000 MJy/sr, while the 160 \mum\ extended
emission varies from 45 -- 400 MJy/sr (with the central region of L1688 
being saturated). This extreme
background variation unfortunately results in highly non-uniform
point source detection sensitivity in our survey. Our region of
worst sensitivity coincides with the central core of L1688
where the most intense past studies have been performed. Thus,
our {\it Spitzer} c2d survey should be considered complementary
to past ISO and ground based IR studies in Oph.

Because a large cloud complex is encompassed by our maps and the
population of the individual clouds are known to have
substantially different characteristics, we have chosen to
separate out the properties of stars in L1688, L1689, L1709, and
the L1712+L1729 filaments; see Figure~\ref{fig:where}. In
addition, we have separated out the properties of the
disconnected northern filaments (L1744+L1755+L1765), which we
identify as Oph North.  The above regions were defined as
rectangles shown in Figure~\ref{fig:where} with corners
given in Table 2.   

\subsection{CSO Sharc II}

Several L1709 sources were observed at 350\,$\mu$m, using Sharc II at the
Caltech Submillimeter Observatory (CSO) on 2007 April 23, 25 and 26th.
Sharc II is a 12$\times$32 bolometer array, with a pixel size of
4.85\asec\ and a field of view of 2.6\amin\ by 0.97\amin. The Dish Surface
Optimization System (DSOS) was used throughout the run, providing a stable
beam, with a FWHM of $\sim$8\asec\ at 350\,$\mu$m.  Sub-mm conditions
were excellent throughout the run, with $\tau_{225}$ ranging from 0.03
to 0.045. Target observations were interspersed with observations of
IRAS 16293-2422 and Callisto, which were used for both pointing and
flux correction.  The data were reduced using the CSO data pipeline
CRUSH, and relative aperture photometry on the sources obtained.  The measurement uncertainties 
include the flux calibration uncertainty, which is typically on the order 
of 20\%.

\section{Results}

\subsection{Multi-wavelength images}
\label{sec:images}

\begin{figure}
        \includegraphics[angle=270, width=15cm]{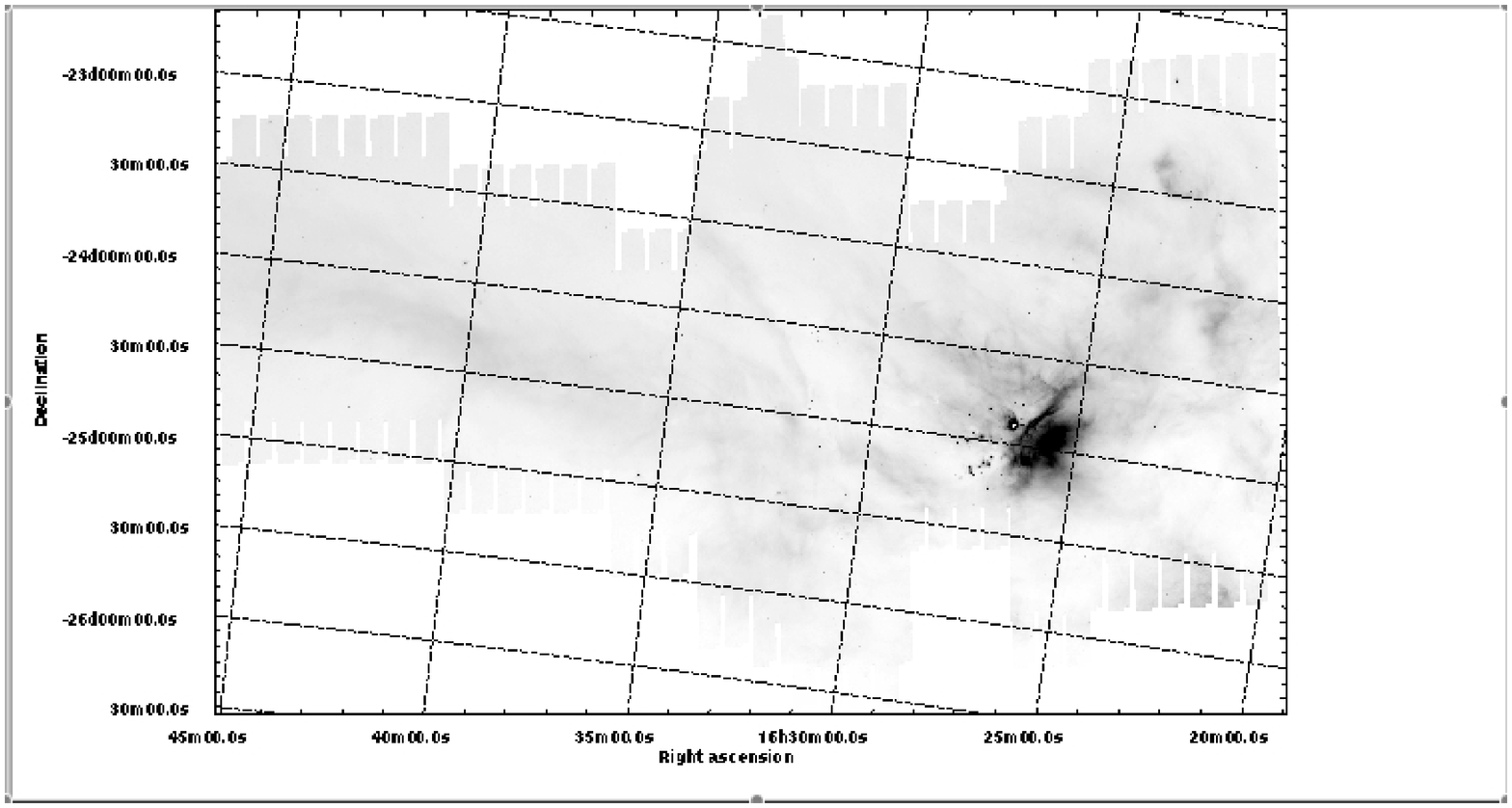}
\epsscale{0.75}
\caption{{\it Spitzer} MIPS 24 \mum\ mosaic of Ophiuchus clouds.   }
\label{fig:24mosaic}
\end{figure}

\begin{figure}
        \includegraphics[angle=0, width=15cm]{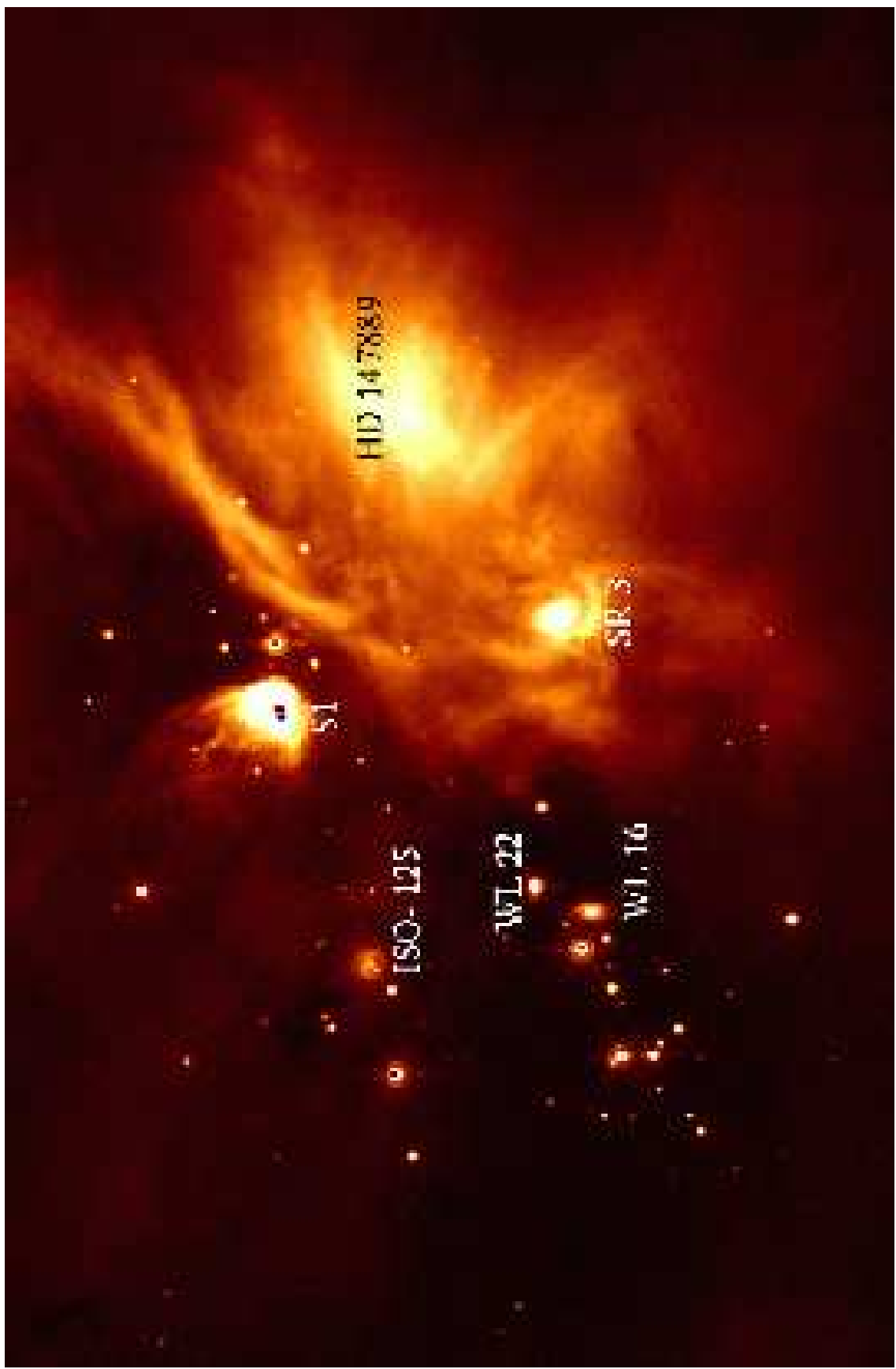}
\caption{MIPS 24 micron image central 1\deg\ x 0.5\deg\ region region 
of Lynds 1688. }
\label{fig:24center}
\end{figure}

\begin{figure}
        \includegraphics[angle=0, width=15cm]{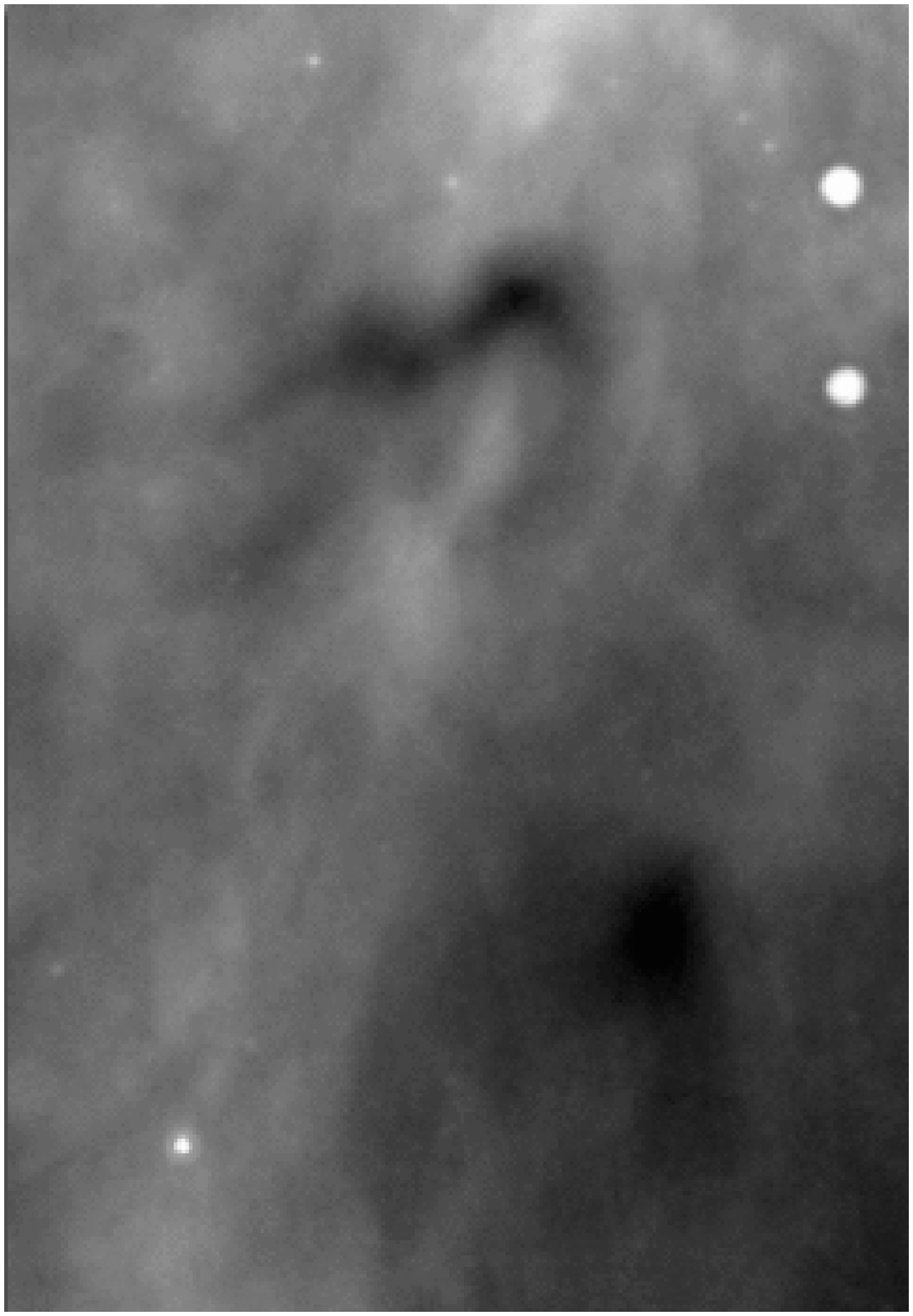}
\caption{24 \mum\ dark clouds found near $16^h28^m31^s 
$-$24^{\circ}$18\amin40\asec.  The image spans 15\amin$\times$10\amin,
with N up and E to the left}
\label{fig:24dark}
\end{figure}

The MIPS 24 \mum\ mosaic (Figure~\ref{fig:24mosaic}) shows
a great deal of diffuse nebulosity and many bright point
sources. The central region of L1688 (Figure~\ref{fig:24center})
is dominated by nebulosity near the early type stars
(S1, SR3) embedded in the cloud.  The S1 source
is saturated in the MIPS 24 \mum\ image, but the peak appears
extended, and it is surrounded by a complex  cometary nebula
opening toward the northeast. The B star SR 3 also appears as a
resolved nebulosity at 24 \mum\ as do the more embedded
early-type objects WL 16 and WL 22.  The WL 22 nebula has a central
core with FWHM of 18\asec and an outer cometary morphology extending
to radii of 30\asec.  The WL 16 nebula is more elongated, 2\amin by 
35\asec along PA 10\deg - 50\deg from the major axis of the PAH 
disk reported by Ressler and Barsony (2003). 
An elliptical ring of emission about 1\amin across is
centered on a point source near the position of ISO-Oph 125 and 
ISO-Oph 124 (YLW 12B and 12A; spectral types F7 and M4 
Greene \& Meyer 1995; Luhman \& Rieke 1999). 
Two other point sources within this ring are also detected at the 
positions of ISO-Oph 128 and ISO-Oph 129. South of these sources, the 
HD 147889 star 
appears only as a faint point source against some of the brightest part
of the central Oph nebulosity and shows a purely photospheric SED. 
This confirms that the large FIR excesses assigned to this source 
by the IRAS PSC are most likely due to interstellar emission, rather
than circumstellar material. Ground based photometry of this 
source (Lada, Wilking, \& Young 1989) also has suggested a lack of
circumstellar material in this system. However, 
HD 148579 (B9V) south of L1689 has
a resolved 24 \mum\ nebula, as does nearby HD 148605 (B3V).
Interestingly, these stars, which are apparently adjacent to or
embedded in cloud material, have not excited the interstellar nebulosity to the
same degree associated with the early type stars in the central
cluster. The 24 \mum\ mosaic also shows a surprising amount of
``dark'' nebulosity (i.e., absorbing perhaps 80\% of the background
illumination, in the vicinity of the L1688 cluster and
extending towards L1689. One of the 24 \mum\ dark cloudlets is
shown in  Figure~\ref{fig:24dark}. 

\begin{figure}
        \includegraphics[angle=270, width=15cm]{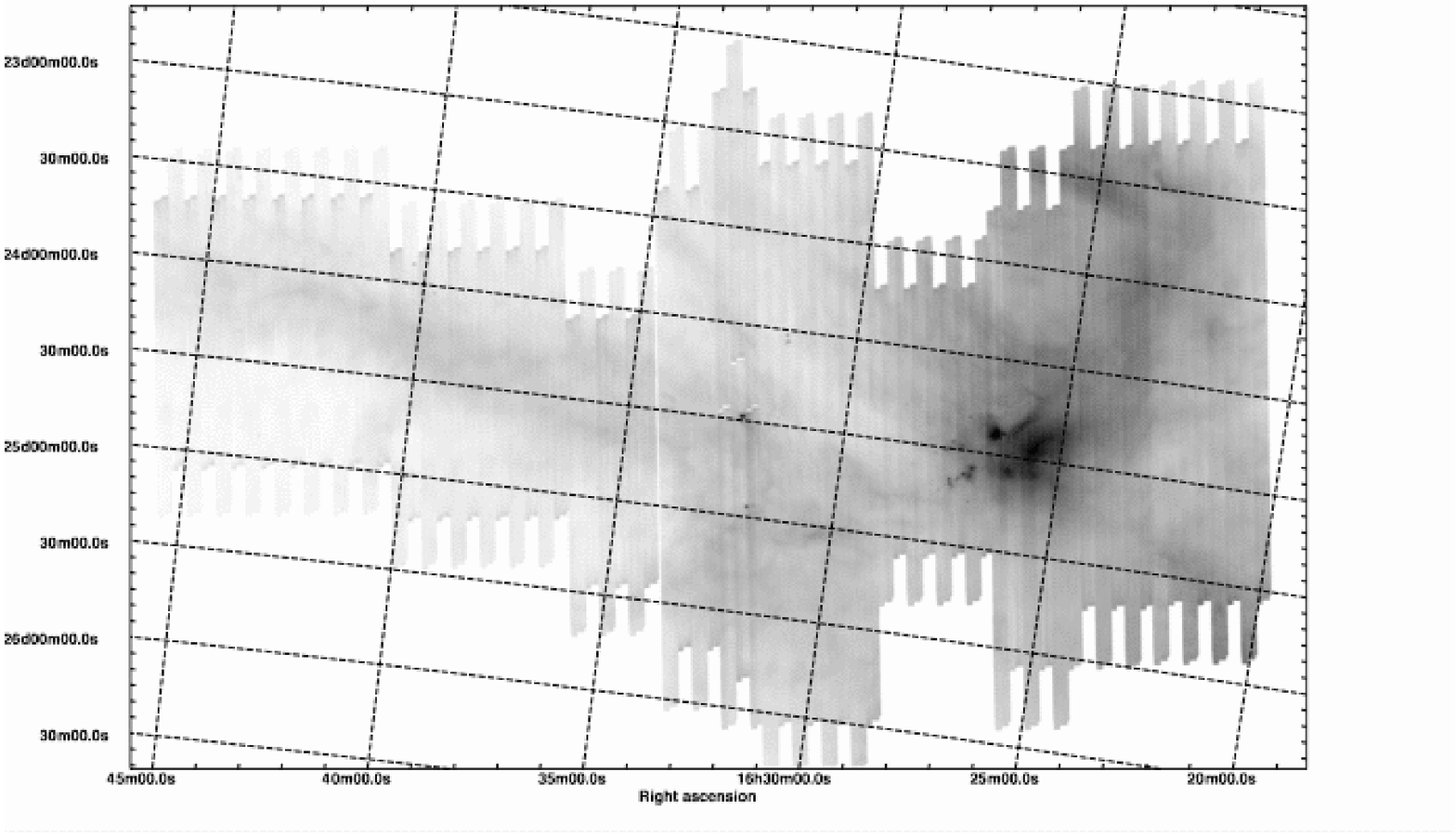}
\caption{{\it Spitzer} MIPS 70 \mum\ mosaic of Ophiuchus clouds. }
\label{fig:70mosaic}
\end{figure}
\begin{figure}
        \includegraphics[angle=270, width=15cm]{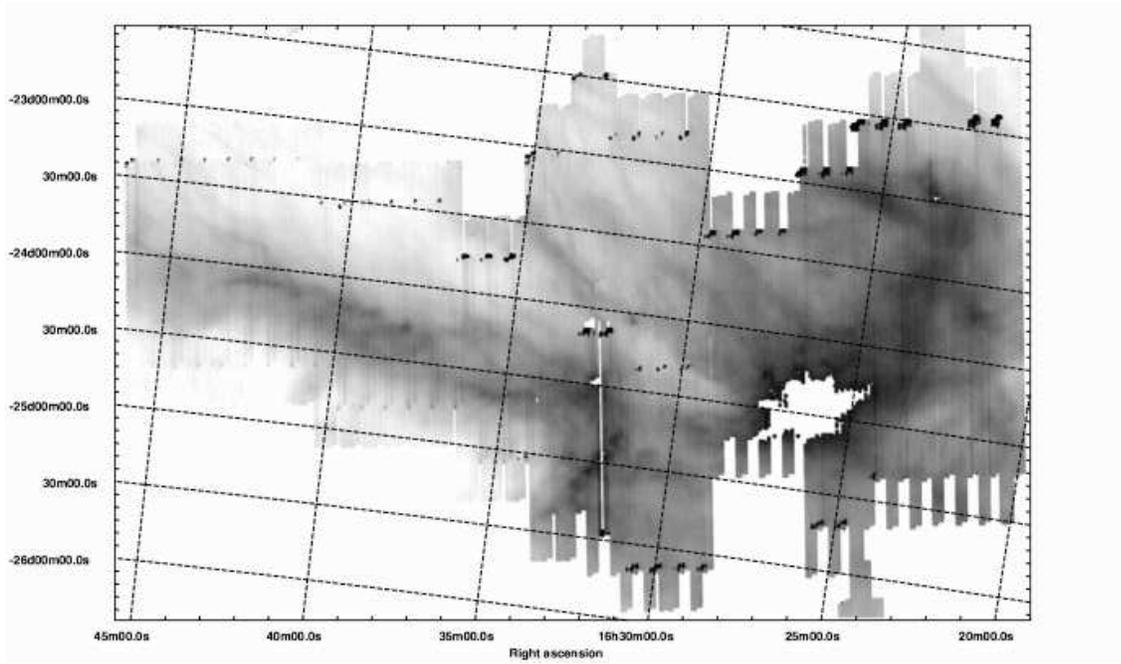}
\caption{{\it Spitzer} MIPS 160 \mum\ mosaic of Ophiuchus clouds. The
central region of the cloud is saturated in our maps.   }
\label{fig:160mosaic}
\end{figure}
\begin{figure}
\epsscale{0.75}
\plotone{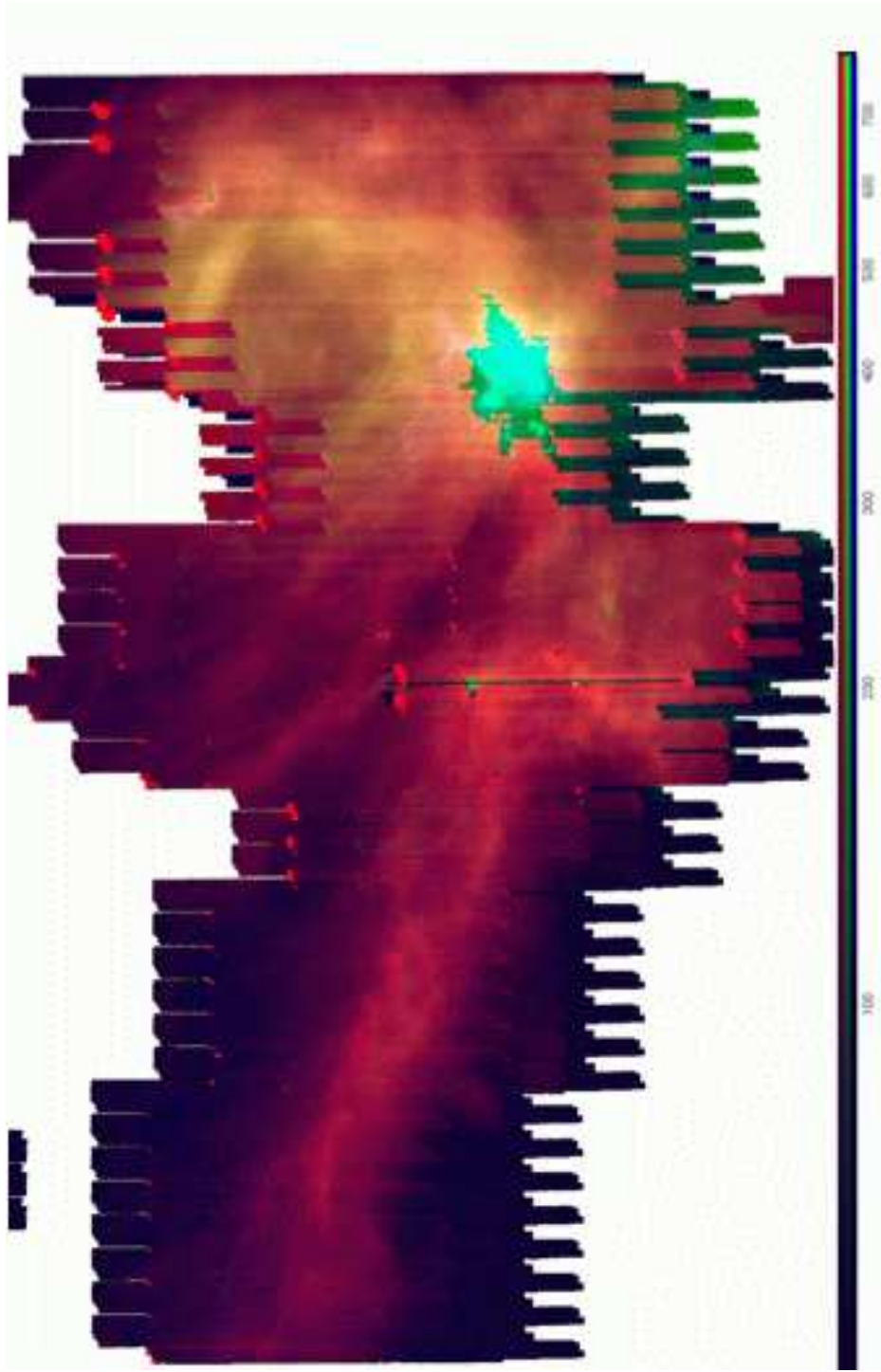}
\caption{24 (blue), 70 (green), and 160 \mum\ (red) mosaic of Ophiuchus clouds.}
\label{fig:3color}
\end{figure}

The longer wavelength MIPS mosaics are shown in 
Figures~\ref{fig:70mosaic} and \ref{fig:160mosaic}. The 70 \mum\
mosaic still shows striations from the detector history effects
which are not entirely removed in the pipeline processing for
each scan leg. Contributing to this appearance is the fact that
the map was filled by interleaving observations at two
different epochs. Despite the cosmetic inperfections, the MIPS
70 \mum\ map clearly shows that the diffuse emission at this
wavelength is brightest in the L1688 region and in the area
extending from northwest to southwest of the central cluster. 
The COMPLETE CO maps of the Oph clouds 
(Ridge et al. 2006) show that this region of Oph West consists of generally 
tenuous cloud material with a few interspersed clumps. It is
possible that these regions are bright in the long-wavelength
MIPS bands because they
on a direct line of sight to the young B stars in the central
cluster and are heated by them. S1, WL 16, WL 22, and YLW 12B
continue to be nebulous at 70 \mum\ as they were at 24 \mum.
There is also a small area of diffuse nebulosity adjacent to 
HD 148605. The sources IRAS 16293-2422 and IRAS
16288-2450 are also prominent at this wavelength. The 160 \mum\
mosaic (Figure~\ref{fig:160mosaic}) shows the same L1688
west dust emission features as the 70 \mum\ image. However, the
cold dust in L1689 and the L1712+1729 filaments becomes
prominent at this wavelength. Young et al.\ (2006) showed that
there are some weak millimeter sources in these filaments. 
The central region of L1688 is saturated in the MIPS 160 \mum\
image. Figure~\ref{fig:3color}  shows a 3-color image with all
three channels of MIPS included. 


\subsection{Source Counts}
\label{sec:sourcecounts}

\begin{figure}
\epsscale{1}
\plotone{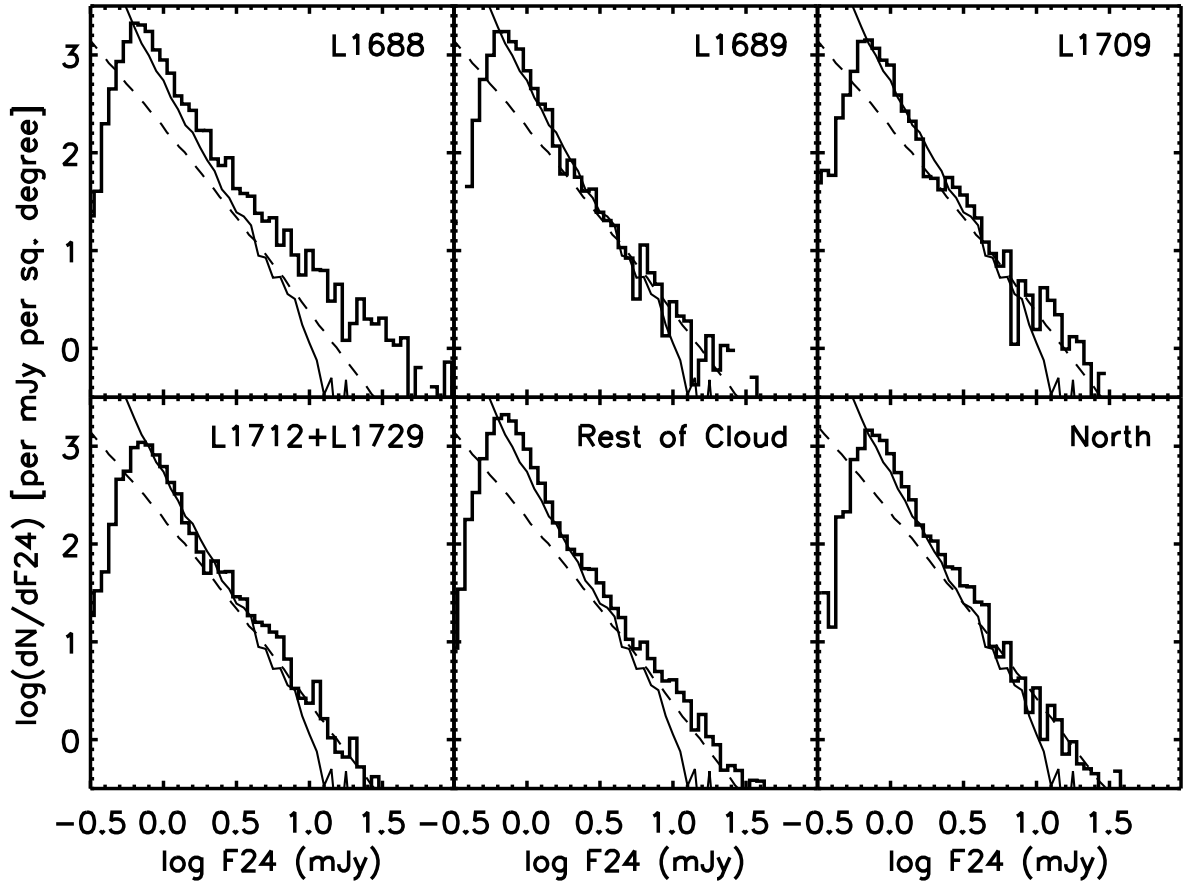}
\caption{Ophiuchus differential number counts at 24 \mum.
The solid line shows the SWIRE ELAIS N1 extragalactic number counts; 
the dashed line shows the Wainscoat model prediction of Galactic star counts.  }
\label{fig:diffnumbercounts24}
\end{figure}

MIPS source counts in Ophiuchus are a combination of cloud
members, foreground/back-ground stars in the galaxy, and the
extragalactic background. An excess of sources above comparison
regions outside the clouds implies the presence of young stars
with disks. Figure~\ref{fig:diffnumbercounts24}
shows the observed Ophiuchus 24 \mum\ differential source counts
in comparison to observed source counts from the SWIRE ELAIS N1
extragalactic field (solid lines), and to the prediction for
Galactic star counts in the IRAS 25 \mum\ band from the
Wainscoat \etal\ (1992) model provided by J.\ Carpenter (2001,
private communication; dashed lines).  
Aside from L1688, all of the other
regions have negligible levels of excess 24 \mum\ source counts,
indicating only small numbers of young stars are present in these 
fields.  L1709 has a modest excess above 10 mJy. These plots indicate
that only L1688 has a substantial young population relative to
the various foreground/background sources expected in the field.

\begin{figure}
\epsscale{1}
\plotone{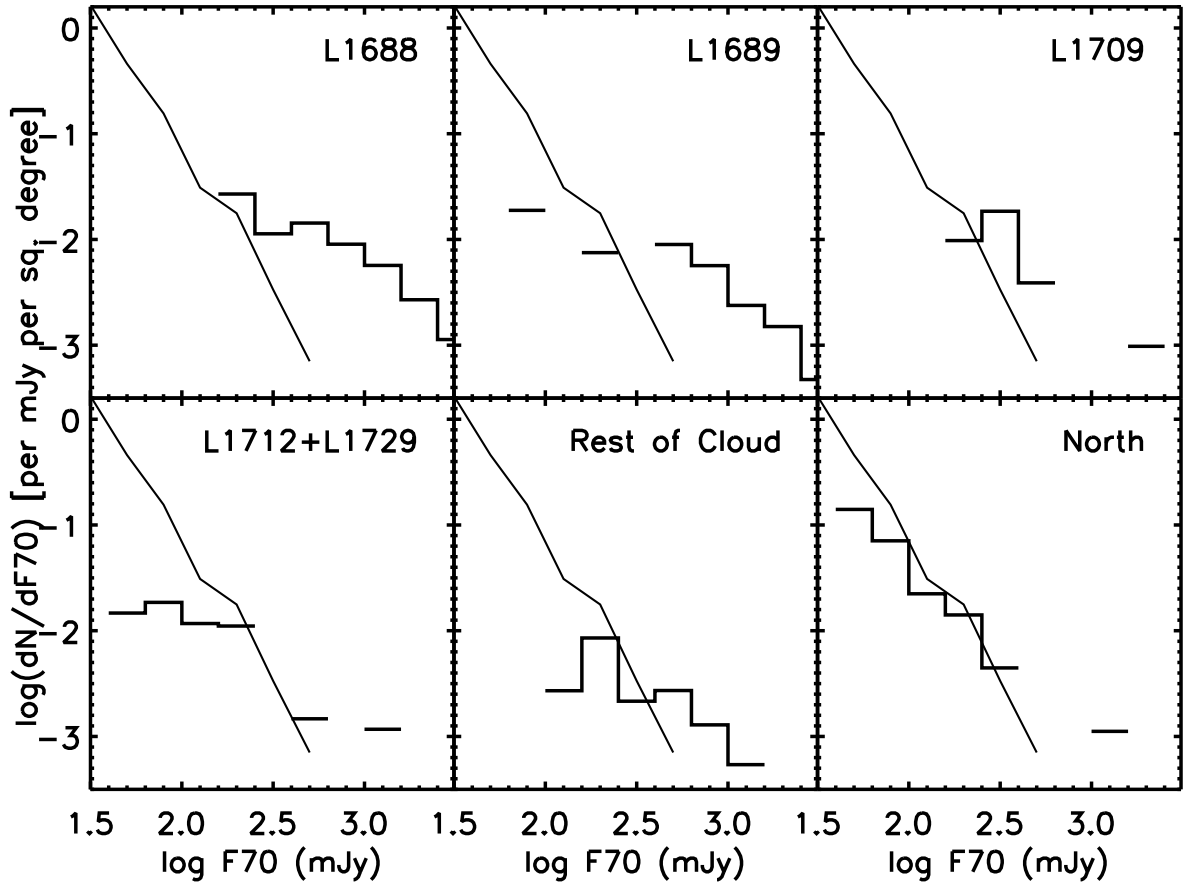}
\caption{Ophiuchus differential number counts at 70 \mum.
The solid line shows the SWIRE ELAIS N1 extragalactic number counts.  }
\label{fig:diffnumbercounts70}
\end{figure}

Figure~\ref{fig:diffnumbercounts70} shows 70 \mum\ source counts
in Ophiuchus, along with extragalactic background counts from
the SWIRE ELAIS N1 field. At the low sensitivity of our
70 \mum\ measurements, detection of normal stellar photospheres 
will be negligibly rare; thus, comparions with Galactic star
counts are not needed. Clearly, given the high backgrounds
and small number of cataloged 70 \mum\ sources, we are dealing
with small number statistics. However, the plots do show
significantly more sources at bright flux density levels for the regions
L1688, L1689, and possibly L1709 than would be found from the
extragalactic background. For L1688, the excess sources extend 
down to 200 mJy, while for the other two regions, the excess above
background 
starts at about 300 mJy. In the extended cloud outside the clusters, 
a slight excess in source counts might be present above a flux 
density of 400 mJy.  Note that the depth of 70 \mum\ extractions 
is greatest in the Oph North field where the diffuse background is 
lowest. In this field, the 70 \mum\ source counts follow the SWIRE 
survey curve to the point where the greater SWIRE sensitivity diverges 
from the c2d sensitivity.

\subsection{Bandmerging with IRAC and 2MASS}
\label{sec:bm}

The IRAC data for the c2d Ophiuchus field (covering $\sim$6.8
square degrees) will be described in Allen \etal\ (2008), and these
data (as bandmerged with the MIPS sources) were included in our
final catalog.  Our MIPS map covers more than twice the area of
the IRAC map.  In the A$_V> 5$ region with full MIPS and IRAC
coverage, there are 2608 MIPS-24 point sources, 88\% of which have an 
counterpart in at least one IRAC band; 31 70 \mum\ point sources, all 
but one of which have an IRAC counterpart; and six 160 \mum\ point
sources, all with IRAC counterparts.  The wide variation in resolution
between IRAC ($\sim$2\arcsec\ at 3.6 \mum) and MIPS (6\arcsec\ at 24
\mum, 18\arcsec\ at 70\mum, 36\arcsec\ at 160 \mum) leads to 
potential confusion of several IRAC objects within a single MIPS 
beam.  Details of the bandmerger are given in Evans et al. (2007)
(c2d Delivery Document).  For the total MIPS map region, 
the above source counts approximately double; see Table 3.
The c2d source catalog containing all Oph {\it Spitzer} sources can be found at:
$http://data.spitzer.caltech.edu/popular/c2d/20051220\_enhanced\_v1/Oph/$

Near-IR $JHK_s$ data from 2MASS (Cutri et al. 2003)
were also used in constructing our
catalog.  The 2MASS data obviously covered the entire region of
the MIPS map, but to a shallower depth than the IRAC
observations, so only 45\% of the MIPS-24 sources have 2MASS $K$
counterparts, similar to the fraction seen in the Cham II
catalog (Young et al.\ 2005).



\subsection{Color-Magnitude and Color-Color Plots}
\label{sec:cmdccd}

For relatively bright sources, K$_s$ versus K$_s-$[24] can be
used to separate potential YSOs from stars and most
extragalactic sources. This metric has been used in a variety
of {\it Spitzer} papers to identify stars with infrared
excess (Gorlova et al.\ 2004, Padgett et al.\ 2004,
Padgett et al.\ 2006, Rebull et al.\ 2007, etc.). 
Stellar photospheres without excess are clustered
around K$_s-$[24] = 0. If K$_s-$[24] $>$ 2, then a strong excess is
present at 24 \mum. Most extragalactic sources are fainter than
K$_s$ = 14. 

\begin{figure}
\epsscale{.80}
\plotone{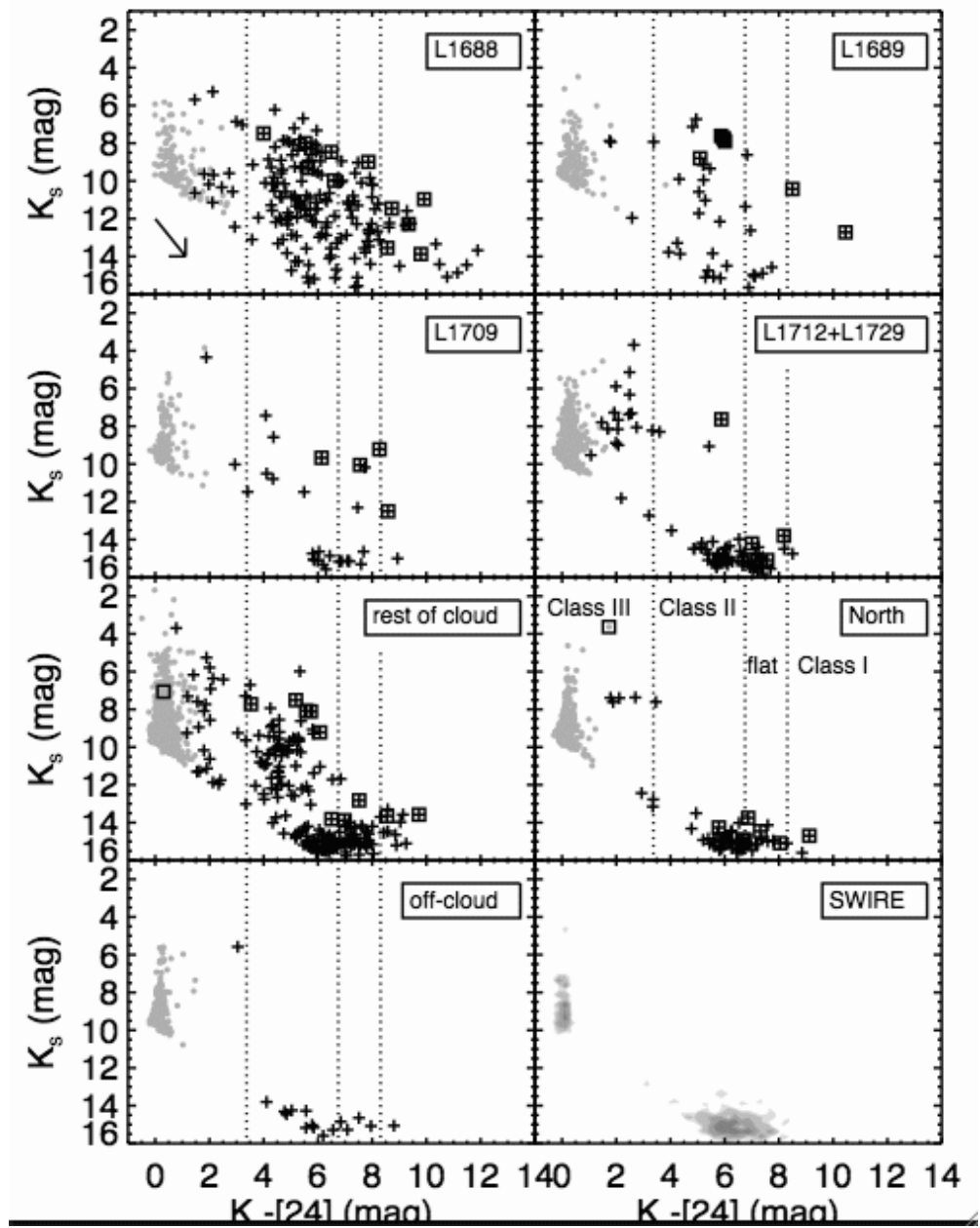}
\caption{$K_s$ vs.\ $K_s-[24]$ for Ophiuchus objects in the various
sub-clusters, as marked, the ``rest of the cloud'', the
off-cloud region, and SWIRE (contour plot, lower right).
Objects in SWIRE are expected to be mostly galaxies (objects
with $K_s\gtrsim$14) or plain photospheres (objects with
$K_s-[24]\lesssim$1; marked as grey on the Oph region
plots).  Boxed sources are also detected at
70 \mum. The vertical dashed lines delineate (from left to
right) regions where Class I, flat-spectrum, Class II, and
Class III infrared source models would reside.}
\label{fig:k_k24}
\end{figure}

Figure~\ref{fig:k_k24} presents K$_s$ versus K$_s-$[24]
color-magnitude diagrams for L1688, L1689, L1709, the
L1712+L1729 filament, Oph North, the rest of the cloud, and the
extragalactic SWIRE ELAIS N1 field. Note that sources that appear to 
be pure photospheres appear in gray.  SWIRE 24 \mum\ sources tend to
congregate in two groups: namely, main sequence photospheres at
color of zero and galaxies at faint K magnitudes and colors of 6
-- 8. Thus, K$_s-$[24] plots help to define regions of
color-magnitude space which are likely to be affected by
extragalactic sources seen through the cloud. 
In Figure~\ref{fig:k_k24} dashed lines (from right to left) denote 
the divisions between Class I, flat, Class II, and Class III objects
following Rebull et al.\ (2007); in brief, we can use the
observed \ks$-$[24] colors and assign an $\alpha$ index
following Greene \etal\ (1994).
Note that the formal Greene \etal\ classification puts no lower
limit on the colors of Class III objects (thereby including
those with SEDs resembling bare stellar photospheres, and
allowing for other criteria to define youth).  In our case,
since we know little about many of these objects, in an
attempt to limit contamination from foreground/background stars
(and background galaxies), we have imposed the additional 
constraint that \ks$-$[24]$>$2, and \ks$<$14.  Note that there
are certainly true young cluster members that do not meet these
(conservative) criteria, but identifying them (and separating
them from the galaxies) is beyond the scope of this paper. Allen 
et al.\ (in preparation) will discuss the combined IRAC and MIPS SEDs and 
present a list of cloud members.

   Interstellar and circumstellar reddening could plausibly affect
the location of sources on the K$_s$ versus K$_s-$[24] diagram.
To test this possibility, we used the extinction law derived
from {\it Spitzer} lines of sight through Ophiuchus (Flaherty
et al. 2007) to produce a reddening vector in Figure~\ref{fig:k_k24}.  
The vector depicts the effects of five magnitudes of K band 
extinction, which moves a star toward the lower right. Note that
unlike the SWIRE sources, which show a well-defined main sequence
at K$_s-$[24] = 0, almost all of the Ophiuchus sources show a
displacement toward redder colors. Given the high extinctions
thoughout this region, many of the apparently small excess sources
in L1712+L1729 and ``rest of cloud" areas may simply be reddened
photospheres. In addition, Class II sources could plausibly be
reddened into the Class I zone. Thus, in regions of very high
extinction, classification of YSOs based on even long wavelength
MIPS data is complicated by reddening effects.

Table 4 provides statistics on the
number of sources classified by K$_s-$[24] as YSOs in the
various subregions of the Ophiuchus map. The Oph North and
off-cloud fields resemble the SWIRE regions, with most
sources clumped in the photospheric and faint galaxy region of
the diagram. The L1712+1729 filament shows a few more bright
photospheres with small excesses than Oph North plus three
potential Class II sourcs. L1709 and L1689 have a higher
frequency of Class IIs than the filament, as well as a few flat
and Class I SEDs. Although the number of sources in L1709 is
small, this region actually has the highest fraction of Class I + 
flat spectrum sources in the region.  These dense clouds also show 
fewer sources in the extragalactic clump at faint K$_s$ and large 
K$_s-$[24] due to their small area and bright 24 \micron\ diffuse 
emission.  Finally, the L1688 region has a high fraction of flat and
Class I SED sources (27\%), and it dominates the region for sheer
number of Class II sources (100).

\begin{figure}
\epsscale{1}
\plotone{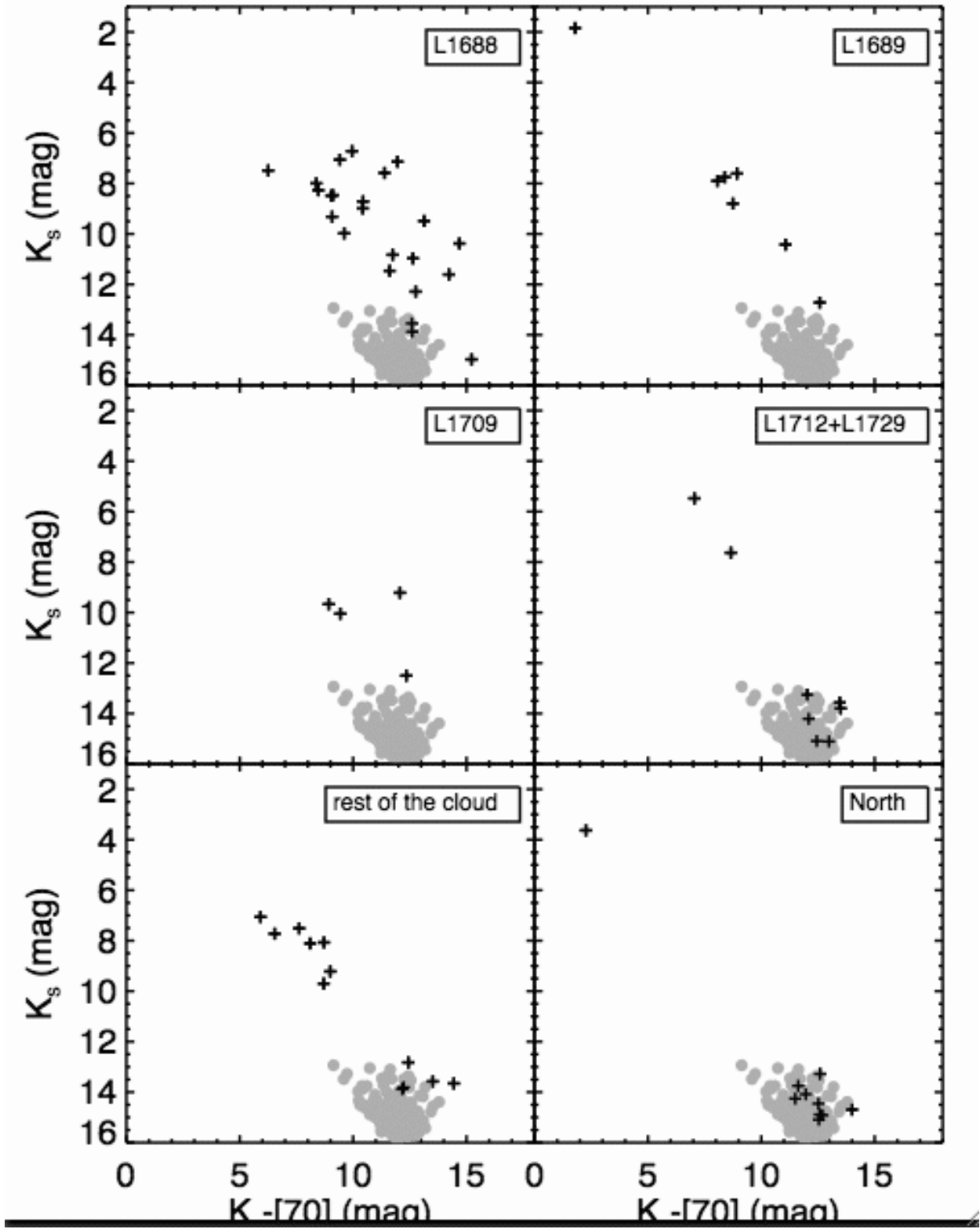}
\caption{$K_s$ vs.\ $K_s-[70]$ color-magnitude diagram for
various regions in Oph (crosses), with data for SWIRE (grey dots) included
for comparison.  See text for details. }
\label{fig:k_k70}
\end{figure}

Figure~\ref{fig:k_k70} shows the K$_s$ versus K$_s-$[70] plots
for Ophiuchus. 
K$_s$-[70] = 0 would indicate a bare stellar photosphere in this plot.  
However, few if any bare stellar photospheres in this field are bright enough
for detection in our very shallow MIPS 70 \mum\ survey. 
Extragalactic SWIRE sources from the ELAIS N1 field are shown
for comparison (grey dots). This plot is most useful for
extremely bright YSOs which are saturated at 24 \mum\, but not
70 \mum. Again, L1688 dominates the number of potential YSOs and
has the most objects with large excesses. These are likely to be
the youngest in the sample. Other regions have only a handful of
70 \mum\ YSOs, and the bulk are in the Class II regime with the
exception of a few sources in L1709 and L1689. There is one
object in L1689 with stellar or near-stellar colors out to 70
\mum; it is SSTc2d 163528.6-245648, which we tentatively associate with
Elias 2-83. This source plausibly may be a reddened star given
the dark cloud visible in this area in the MIPS 24 \mum\ image. 
All the 70 \mum\ sources in Oph North are
consistent with extragalactic background sources, except for
SSTc2d 164417.8-220648, which can be identified with IRAS 16413-2201,
a known carbon star NC 80 (Guglielmo et al.\ 1993). 

\begin{figure}
\epsscale{.80}
\plotone{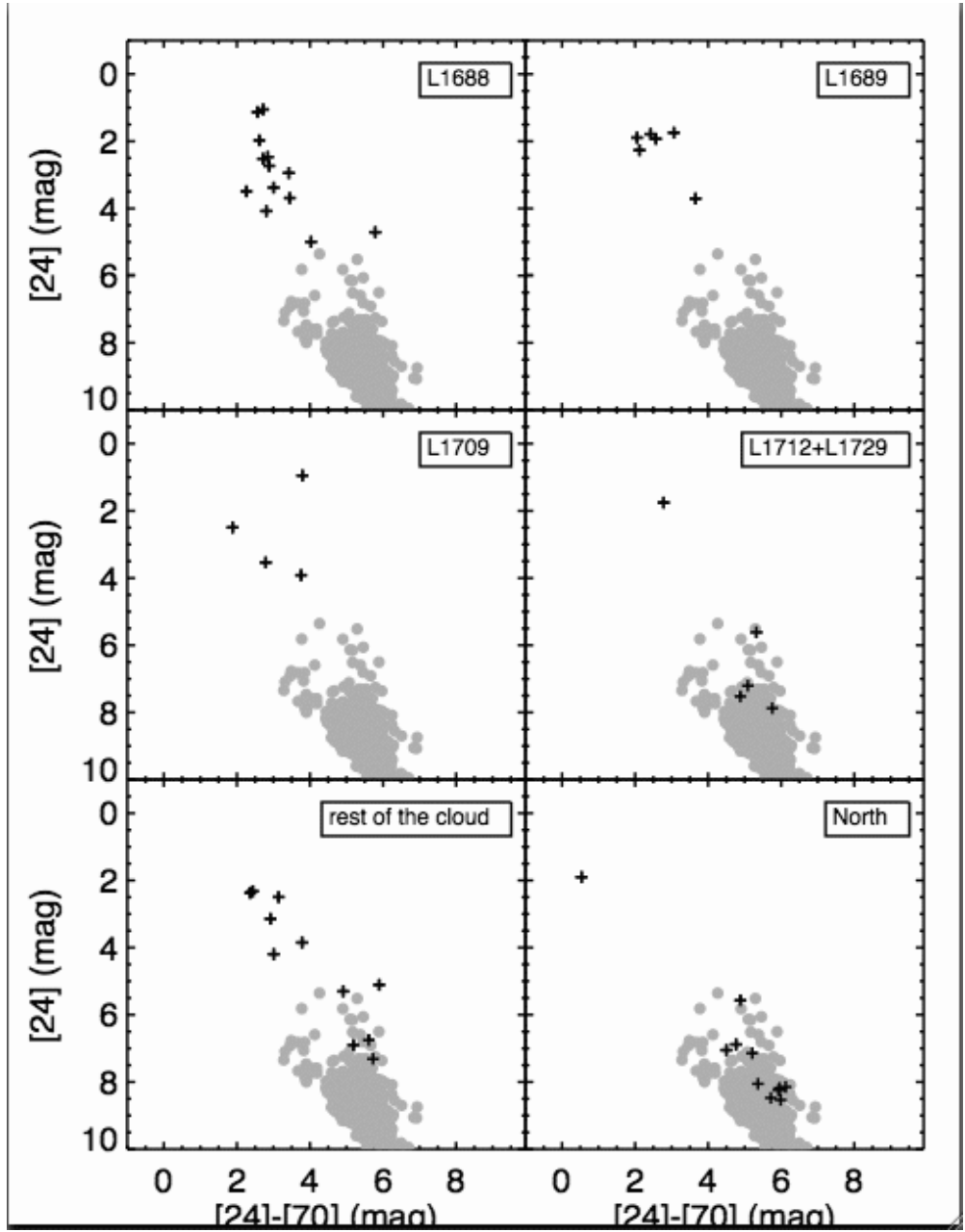}
\caption{[24] vs.\ [24]$-$[70] color-magnitude diagram for
Ophiuchus (crosses), with data for SWIRE (grey dots) included for
comparison.  As in Figure~\ref{fig:k_k70}, the two clumps
correspond to extragalactic ([24]$\sim$7-10) and YSO candidates
([24]$\lesssim$5), and there are few extragalactic objects in
the clusters because the high nebulosity limits the depth of the
survey.  Very red sources ([24]$-$[70]$>$6) are the most
embedded objects.  }
\label{fig:24_70}
\end{figure}

Figure ~\ref{fig:24_70} presents a plot of [24] versus [24] $-$
[70] for the Ophiuchus clouds. In the unlikely case of a bare
stellar photosphere, [24] $-$ [70] = 0. Again, SWIRE sources are depicted
as grey dots. Sources with large infrared excesses will have
large [24]-[70] colors. This color-magnitude diagram shows the separation 
of YSOs from extragalactic sources and
emphasizes that extragalactic 70 \mum\ sources are only seen in
the lower background regions. Although the number of sources
seen in these plots is considerably smaller than in the 24 \mum\
only plots (due to the lower sensitivity of the MIPS 70 \mum\
channel), this color-magnitude diagram suggests confirmation of
the picture of the L1688 cluster having a lower fraction of
very young sources than L1709. However, it is true that a fair
number of the most luminous MIPS sources in L1688 are saturated
at 24 microns ($\geq$ 4 Jy), thus eliminating them from these plots.  
In addition, the small number of sources in L1709 makes the
comparison of questionable significance.  Interestingly, some low 
luminosity sources with large fractional excesses appear to be 
found outside of the described subclusters (in the ``Rest of 
the cloud'' areas).  
Note that 
the brightest objects in these plots (those with [24]$\leq$5 and 
[24]$-$[70]$\leq$6) appear for reference in Table 7 below.

\begin{figure}
\epsscale{0.5}
\plotone{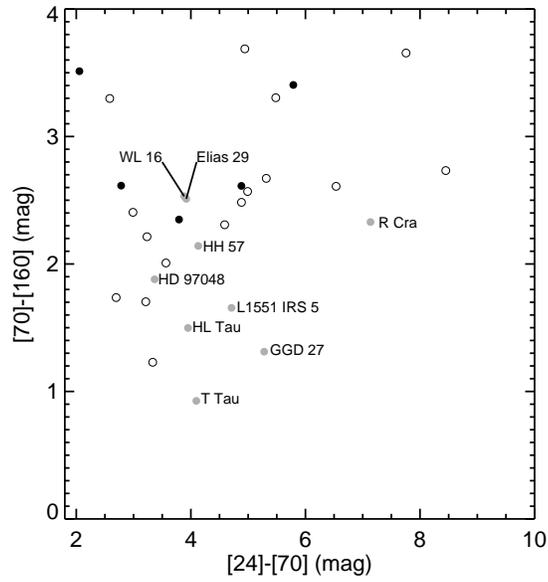}
\caption{Color-color diagram for sources
detected but not saturated in all three MIPS bands.
Objects shown are from the current Ophiuchus survey
(filled circles), the c2d MIPS Perseus survey
(open circles; Rebull et al. 2007),
and bright objects studied by ISO
(grey dots, Noriega-Crespo 2005, private communication).   }
\label{fig:2470160}
\end{figure}

Unfortunately, due to very high backgrounds and saturation
issues, as well as the difference in sensitivity
between the various MIPS arrays, very few point sources (5) 
have measurable detections at all three 
MIPS bands. Figure~\ref{fig:2470160} presents a plot  of
[70]$-$[160] versus [24]$-$[70]. Included for comparison on this
plot are colors calculated for MIPS bands from ISO SWS data for
a variety of very famous well-studied embedded objects from A.\
Noriega-Crespo (private communication);  to
facilitate comparison, these are the same
objects used in a similar plot in Rebull \etal\ (2007).
The four Ophiuchus sources all lie
between [70]$-$[160] of 2.5 to 3.5 while spanning a range of 4
magnitudes in [24]$-$[70]. This may be an extinction effect,
since three of the five objects are located in 
regions of Oph with apparently high extinction even at 24 \mum.


\subsection{Comparison to IRAS}
\label{sec:iras}

As in Rebull \etal\ (2007) for Perseus, the MIPS data for Oph
offer an opportunity to assess how successful the IRAS survey was
in identifying point sources in a complex region. We now
explicitly compare the IRAS Point Source Catalog (PSC; Beichman 
\etal\ 1988) and IRAS Faint Source Catalog (FSC; Moshir \etal\ 
1992) results in this region to the Spitzer
c2d images and catalogs described in this paper.  As for
Perseus, extended emission is present in all three MIPS bands,
posing a significant source of confusion to the large aperture
IRAS measurements.  Many of the IRAS PSC objects were
detected at 60 or 100 \mum\ only, with only upper limits at 12
and 25 \mum; so even without Spitzer data, one might suspect
that such sources might correspond to texture in the extended
emission.  As for Perseus (Rebull \etal\ 2007), the Spitzer data
comparison clearly shows that significantly fewer spurious
sources appear in the FSC than in the PSC.

A summary of the comparison of MIPS sources to IRAS sources are
found in Table 5. For each IRAS PSC and FSC band, the number
of IRAS detections are listed, followed by the number of
these IRAS detections recovered by the MIPS 24 and 70 \mum\
arrays, IRAS detections not found by MIPS, IRAS detections determined 
to be nebulosity by MIPS, and IRAS detections
resolved as multiple by MIPS. It is not surprising
that few of the long wavelength IRAS sources are recovered by
MIPS 24 \mum. About 65\% of the 25 \mum\ PSC
sources are recovered as point sources in MIPS-24; only 23\% of
the 60 \mum\ PSC sources are recovered at MIPS-70. The most
frequent reason for sources not being retrieved is confusion by
nebulosity; 23\% of the 25 \mum\ PSC sources and 57\% of the 60
\mum\ PSC sources are so affected. Surprisingly few objects are
resolved into multiple sources by Spitzer.  Only one 25 \mum\
PSC detection is completely missing from MIPS-24; 16 of the 60
\mum\ PSC detections are completely missing from MIPS-70.  A
list of IRAS PSC sources not cleanly recovered appears in
Table 6.

The FSC does a much better job of finding short-wavelength point
sources, and is {\it much} less confused by the nebulosity.
However, the FSC does not contain any sources on the West side
of the cloud due to high surface brightness in the IRAS images.
All but one of the 25 \mum\ FSC sources are recovered by Spitzer
at 24 $\mu$m; the remaining source is resolved as multiple sources when
viewed by MIPS. There are no 60 \mum\ FSC high quality point
sources in this region.

While it would be interesting to study long-term far-infrared source 
variability in Oph, this is very difficult in practice.  The vastly larger 
IRAS beam means that point source flux densities can include large contributions
from bright background nebulosity.  The different calibration approaches 
of the two instruments requires that any flux density comparison be made
only after large color corrections are applied - corrections that can
only be made accurately if the spectral slope is known across the photometric
bandpasses.  In addition, many sources detected by IRAS at 25 \mum\ are 
near the upper end of the MIPS 24 \mum\ dynamic range, where instrument
calibration is less secure.  For these reasons, we make no attempt to 
compare the MIPS and IRAS photometry for the sources detected in both 
surveys.

\subsection{Time variability}
\label{sec:variability}

\begin{figure}
\epsscale{0.5}
\plotone{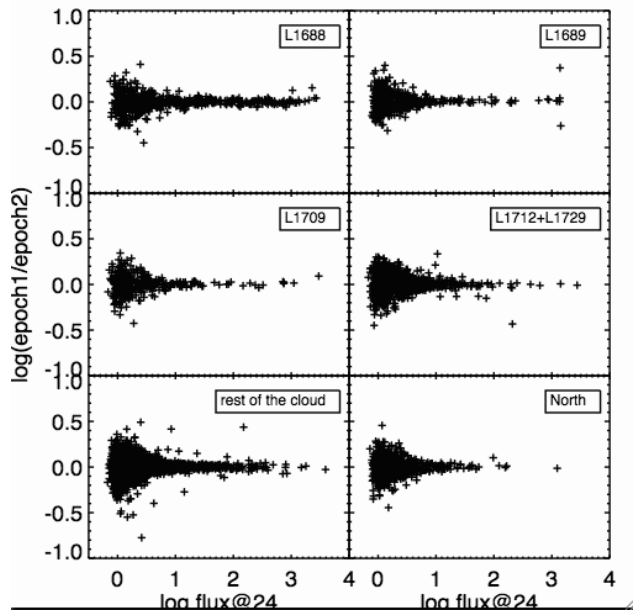}
\caption{ Variability of Ophiuchus 24 \mum\ sources between
two observations separated by 3 -- 8 hours. All cases
of potential variability are well explained by instrumental
effects } 
\label{fig:var}
\end{figure}

Because we have two epochs of observation at 24 \mum, we can
look for time variability on timescales of a few hours. 
Figure~\ref{fig:var} shows the ratio of flux densities determined in the
two epochs as a function of mean 24 \mum\ flux densities.  It can be seen
that typical variability is $\sim$10\%, consistent with our
reported measurement uncertainty.  There are four points whose
flux ratio (in the log) is greater than 0.45, and two whose flux
ratio is less than $-$0.5.  Out of these six sources, three fall on 
the edge of the map, and the apparent variability is clearly due 
entirely to the low redundancy and S/N at the ends of scan legs. 
Three additional objects fall in the L1688 region, where bright sources
cause the readout dependent gain variation known as ``jailbars.'' 
Two of the L1688 objects are faint and located right on top of some 
strong jailbars, and one is deep within the bright, extended emission.  
Therefore, apparent variability in all cases is entirely due to 
instrumental effects. 

\section{Discussion}

\subsection{Individual Objects}
\label{sec:indobj}

\subsubsection{Young Stellar Objects with Extreme K$_s$-[24] Colors}

\begin{figure}
\epsscale{0.75}
\plotone{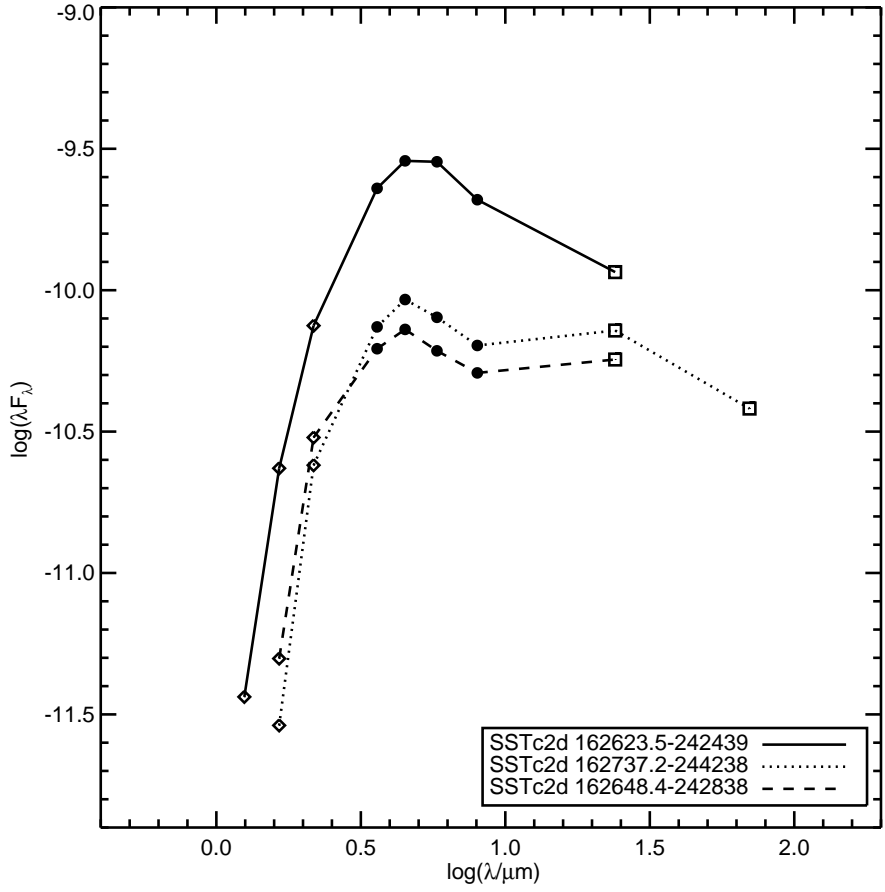}
\caption{
Infrared spectral energy distributions of several MIPS 
sources with K$_s-$[24] $>$ 8. }
\label{fig:sed}
\end{figure}

YSO modelling indicates that some of the earliest evolutionary
states of star/disk systems will be detected by {\it Spitzer}
as point sources with extremely red colors (Robitaille et al. 2006). 
A search through our catalog for sources with K$_s<$ 12 and
K$_s$-[24] $>$ 8  yielded eight sources Their SEDs are
presented in Figure~\ref{fig:sed}. One of these objects (SSTc2d
162623.5-242439) had been previously identified as CRBR
2322.1-1754 (Comeron et al.\ 1993). In that paper, the source
was detected out to Q band, but was not characterized further. In
our data, this source is probably confused at 24 \mum\ with GY21, a flat
spectrum YSO a few arcsec away (Greene et al. 1994).
Its SED is characteristic of an extincted stellar photosphere with 
an excess which appears to be present in the IRAC bands and extends 
to 24 \micron. No 70 \micron\ source is detected. It lies close to
another YSO (Oph S2) (best seen in our 24 \micron\ image), and thus
there is potential SED confusion at longer wavelengths.
Source SSTc2d 162648.4-242838 is
similar, although fainter. It also appears as a binary source in
the IRAC 3.6 \micron\ band and can tentatively be identified with
WL 2 (Wilking et al. 1989). Finally, SSTc2d 162737.2-244238 (GY 301)
is a faint stellar source with an almost flat spectrum out to 70
\micron. All of these sources appear in high extinction regions,
and it appears likely that their extraordinary K$_s-$[24] color
is primarily due to extinction at the K$_s$ band. 
Thus, these are likely normal flat spectrum YSOs with disks seen 
at very high extinctions due to location within the cloud or envelope.

\subsubsection{The L1709 Aggregate and a Possible Edge-on Disk}

\begin{figure}
\epsscale{0.75}
\plotone{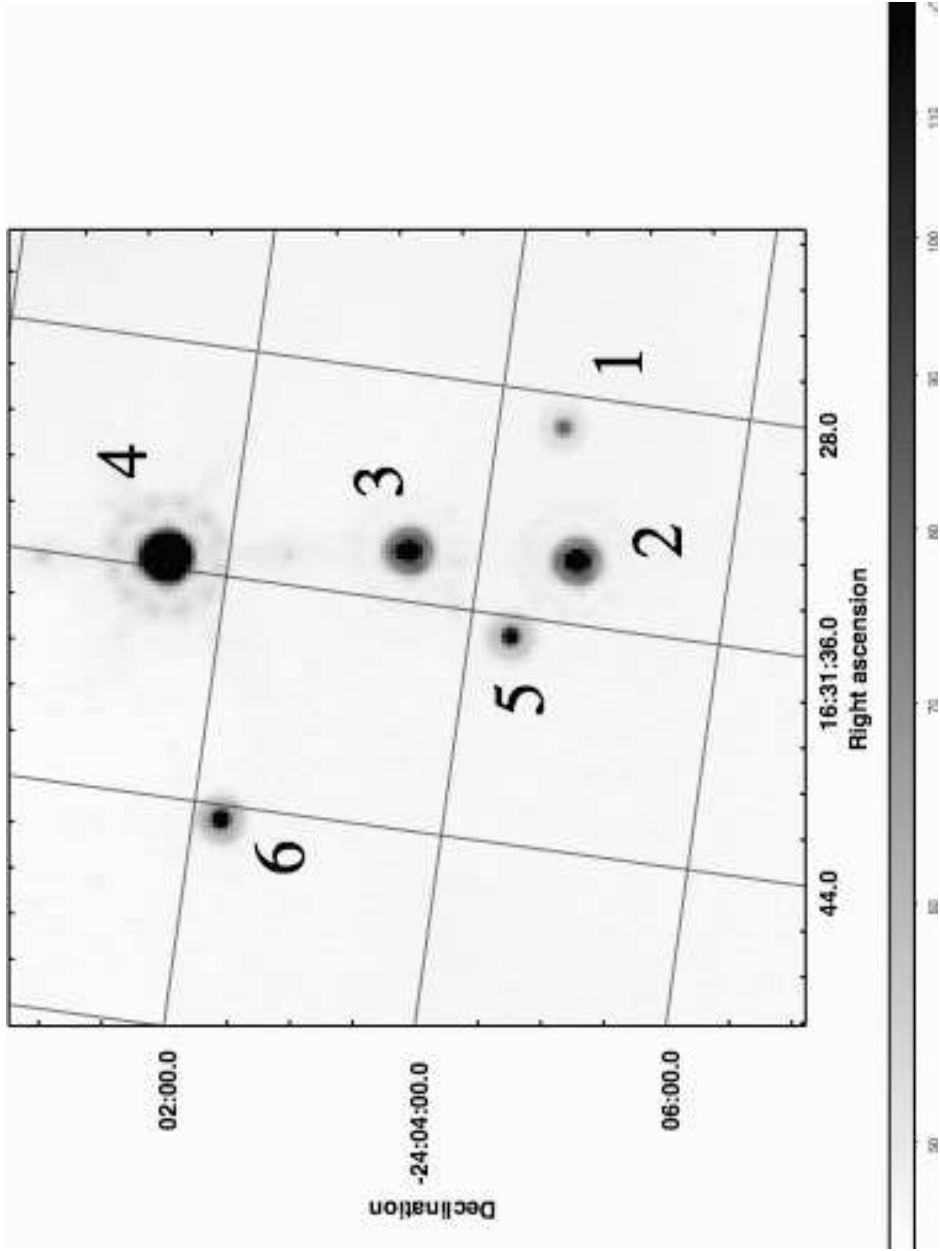}
\caption{24 \mum\ image and finder chart for the L1709 aggregate. } 
\label{fig:1709finder}
\end{figure}
\begin{figure}
\epsscale{0.75}
\plotone{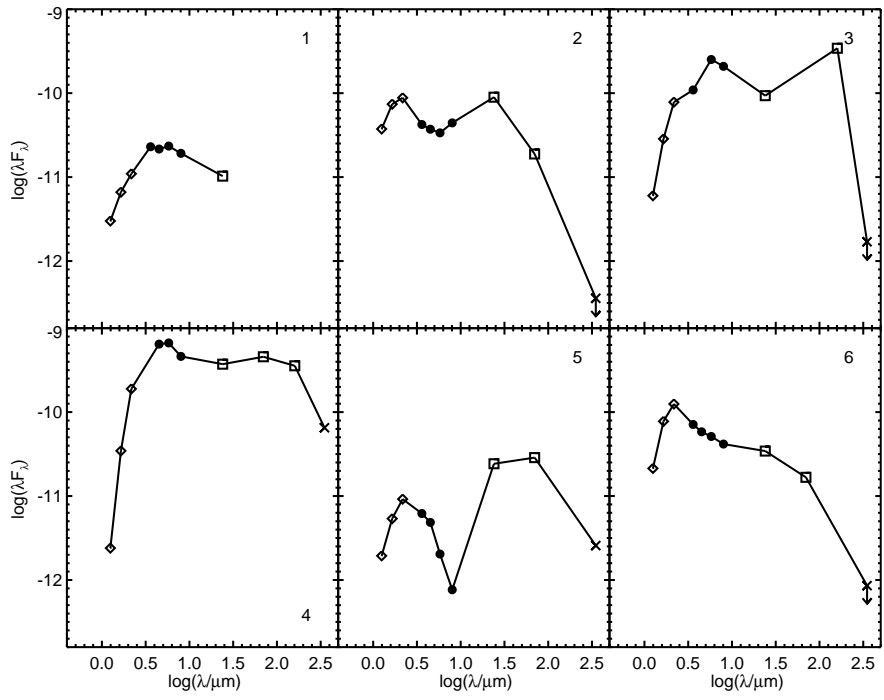}
\caption{SEDs for six stars bright at 24 \mum\ found in the 
the newly identified L1709 aggregate. }
\label{fig:1709agg}
\end{figure}
\begin{figure}
\plotone{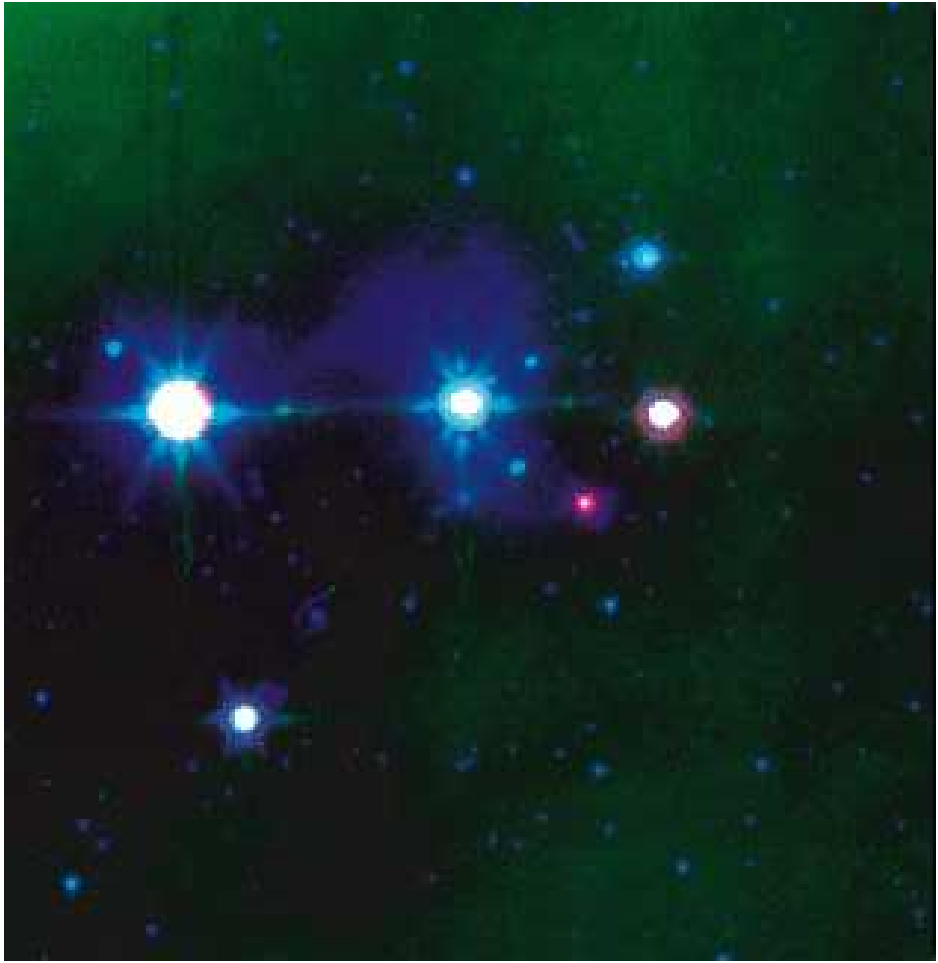}
\caption{
Three color 9\arcmin\ x 9 \arcmin\ image of the L1709 aggregate.
North is up, and east is left. The IRAC 4.5 $\mu$m channel is
shown in blue, IRAC 8 $\mu$m green, and MIPS 24 $\mu$m in red.
The candidate edge-on disk (source 5) is clearly extended in
the 4.5 \mum\ image.}
\label{fig:eod}
\end{figure}

In L1709, there is a grouping of six bright 24 \mum\ point
sources within $4.2\arcmin\times3.3\arcmin$ (see Figure~\ref{fig:1709finder}).
SEDs for all six
of these objects appear in Figure~\ref{fig:1709agg} and MIPS
fluxes for these objects appear in Table 7.  As a shorthand, numbers 
between 1 and 6 appear in Figure~\ref{fig:1709agg}; the corresponding 
formal c2d names appear in Table 8.
IRAS 16285-2358 can be identified with source 2 in
Figure~\ref{fig:1709agg}.  This object was selected as a
candidate YSO by Beichman \etal\ (1986).  
IRAS 16285-2355, source 4 in Figure~\ref{fig:1709agg}, is also
selected as one of the most extreme \ks$-$[24] colors in the
cloud. This well-known source (e.g., Visser et al.\ 2002 
and references therein) shows a typical Class I or flat spectrum
SED rising through the mid-IR and flattening across the three
MIPS bands. It appears to be a higher luminosity version or more
pole-on disk geometry of the same type of sources with extreme
\ks$-$[24] colors discussed in the previous section.
Objects 1, 3, 5, and 6 in Figure~\ref{fig:1709agg} do not have any
known counterparts in the literature.  Source SSTc2d
163136.7-240420, source 5 in Figure~\ref{fig:1709agg}, stands
out for its similarity to model edge-on disks similar to
such objects as HH 30 (Burrows et al.\ 1996).  The 2MASS+IRAC
part of the SED resembles  a normal stellar photosphere which
peaks at K, albeit a very faint one. However, its 24 \mum\ and
70 \mum\ fluxes are almost two orders of magnitude higher than
the 8.0 \mum\ value. Edge-on disks such as HH 30
and HK Tau/c (Wood et al.\ 2002) can appear in scattered light
until 10 -- 20 microns, suppressing both their short and long
wavelength flux densities. A 3.6 \mum\  image is presented in
Figure~\ref{fig:eod}.

Five of the six sources in L1709 were observed in the 350 $\mu$m continuum
at the CSO. Two (L1709-3 and 4) were clearly detected and showed 
extended structure.
For these sources, an elliptical aperture with radius of 1\amin 
used to calculate the flux density, excluding some of the low level 
extended structure in the images.  The L1709-5 (SSTc2d163136.7) source is 
in a region of considerable structure - it appears to be sitting on a 
ridge of emission and shows a cometary structure extending toward the 
south-south-east. To the north-west ($\alpha$=16 31 35.33, 
$\delta$=-24 03 46.8) is another bowed extended ridge of emission, 
with a flux of $\sim$200 mJy. The CSO flux densities and upper limits
are incorporated in the SEDs presented in Figure~\ref{fig:1709agg}. 
We use the equations of Beckwith et al.
(1990) to estimate lower limits on the circumstellar mass from these 
measurements. Presuming that the 350 micron emission is 
optically thin and T$_d$ = 100 K, $\kappa_{\nu}$ = 0.071, 
leading to lower limits on the
circumstellar masses of 1.6$\times10^{-2}$ \msun\ for SSTc2d163135.6-240129 
(L1709-4) and 6.4$\times10^{-4}$ \msun\ for SSTc2d163136.7-240420
(L1709-5).


\subsubsection{L1689}

\begin{figure}
\plotone{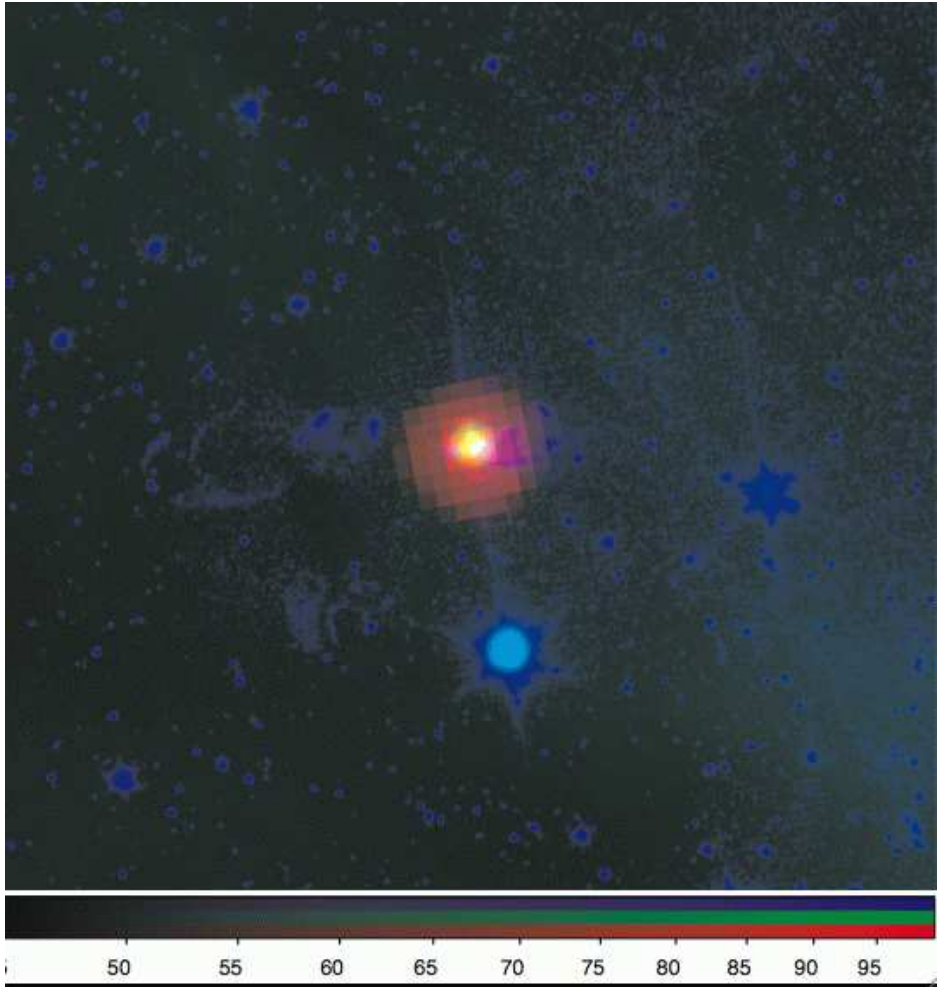}
\caption{
Three color 9\arcmin\ x 9 \arcmin\ image of IRAS 16293-2422.
The IRAC 8 $\mu$m channel is
shown in blue, MIPS 24 $\mu$m green, and MIPS 70 $\mu$m in red.
The source is invisible at 8 \mum, although the outflow emission and
reflection nebulae are visible. Note especially the bow shock features
in blue to the left and right of the central source. In this
image, north is up, and east is
to the left. } \label{fig:16293}
\end{figure}

The famous Class 0 source IRAS 16293-2422 appears in the L1689
part of our MIPS maps. Interestingly, the source is entirely
invisible in the IRAC 8 \mum\ image, although some of the jet
and bowshock emission from the outflow is plainly visible. The
source itself begins to be seen at faint levels at 24 \mum, is
very bright at 70 \mum, and is saturated at 160 \mum. A color
image is shown in Figure~\ref{fig:16293}. The 8 \mum\ visibility of
shocked gas very close to the invisible central source implies
that the high extinction region is very confined in areal
coverage, suggesting an edge-on orientation to the disk/envelope
system. This source has also been
observed by the {\it Spitzer} Infrared Spectrograph (IRS)
(J{\o}rgensen et al.\ 2005). In this work, the authors
interpret the rapid rise in the SED as a large inner hole in an
extended envelope around a protobinary. We prefer to interpret the
very large slope in the SED as an extinction effect from a highly
inclined disk/envelope system.

\begin{figure}
\plotone{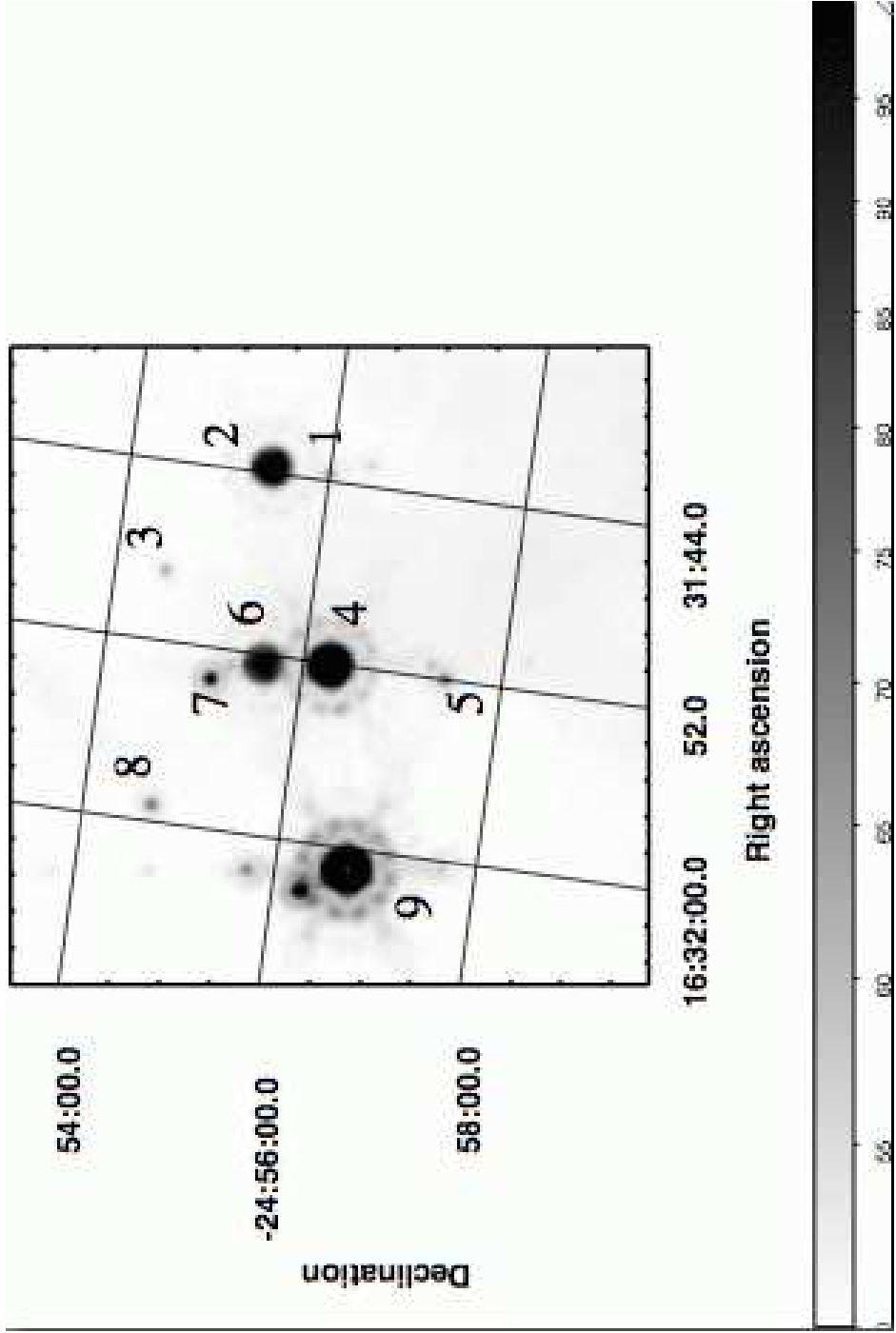}
\caption{24 \mum\ image and finder chart for the L1689 
aggregate. } \label{fig:l1689agg}
\end{figure}

\begin{figure}
\epsscale{0.75}
\plotone{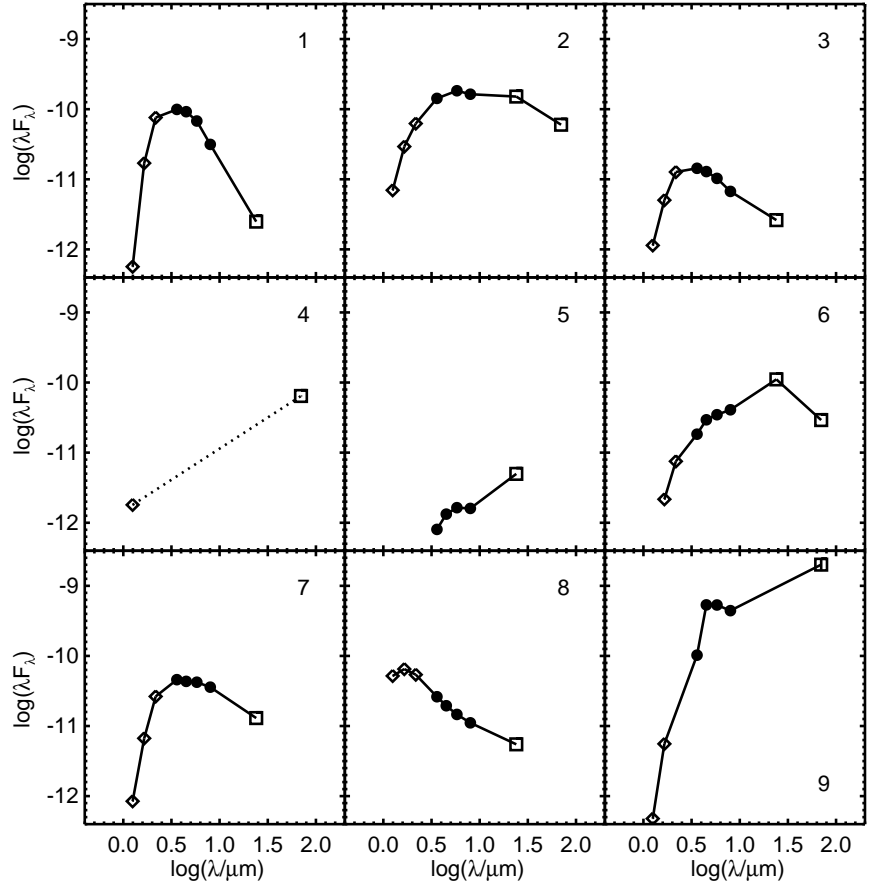}
\caption{SEDs for nine young stellar object candidates
bright at 24 \mum\ found tightly clumped
together in L1689. Source 4 is saturated in all {\it Spitzer} bands
except 70 \mum\ }
\label{fig:l1689sed}
\end{figure}

There is an aggregate of nine stars bright at 24 \mum\ found clustered
within a 5$\arcmin$ region within L1689.  The MIPS-24 image of the
aggregate is presented in Figure~\ref{fig:l1689agg}, and the 
{\it Spitzer}er+2MASS SEDs for these stars appear in Figure~\ref{fig:l1689sed}.
Several of these objects were also seen by ISO and discussed
as part of Bontemps \etal\ (2001).  



Padgett \etal\ (2004) and Rebull \etal\ (2007) commented upon small
clusterings of sources bright at 24 found in L1228 and Perseus,
respectively.  While Oph does not have such aggregates in as high
abundance as in Perseus, this grouping in L1689 is clearly a similar
phenomenon in that there are several sources within $\sim$0.15 pc, all
bright at 24 \mum\ and with a diversity of SEDs.  As can be seen in
Figure~\ref{fig:l1689sed}, some sources resemble reddened photospheres
with just a hint of excess at 24 \mum\
(e.g., \#1), some appear to have Class II SEDs (e.g., \#8), and some have
substantial excesses like Class I or flat spectrum SEDs (e.g., \#6).  
Source \#5 has a strangely shaped
SED because it is an unresolved binary at 24 \mum.  Source \#9 is so
bright at 24 \mum\ as to leave several latents on the array (e.g., it
is near saturation), so the unusual shape of the IRAC portion of the
SED is likely to be a result of saturation in those bands. Object \#4
is saturated at all Spitzer bands except for 70 microns.  The
proximity of these sources suggest that they are physically
associated, despite their disparate SEDs. As in L1228 (Padgett et al.
2004), this aggregate seems to have sources at a variety of stages
in their circumstellar evolution although the sources must be nearly
coeval. 


\subsection{Clustering of YSOs in Ophiuchus}
\label{sec:cluster}

\begin{figure}
\epsscale{0.75}
\plotone{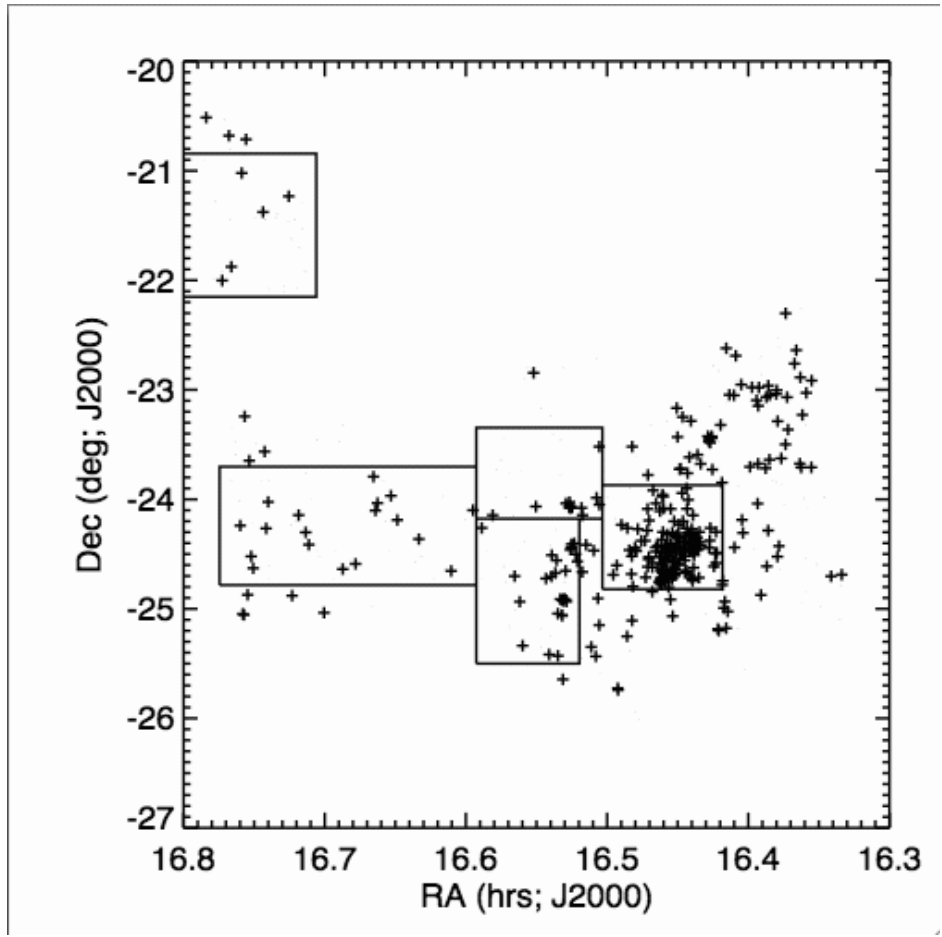}
\caption{Locations of sources where $K_s-[24]>2$ and $K<14$. }
\label{fig:where24x}
\end{figure}

Figure~\ref{fig:where} shows the distribution of MIPS 24 \mum\
sources in the c2d maps. The sources appear to be randomly
distributed in most of the map.  However, as expected, there is
a large overdensity of sources in the region around the Oph A
core. Figure~\ref{fig:where24x} accentuates this result, showing
that relatively bright sources with 24 \mum\ excess are strongly
concentrated in the core, as well as a scattering of small
groups in L1689 and  northern L1688 and southern L1709. A modest
number of excess sources are more widely distributed over the
entirety of the L1719+1729 eastern filament, while Oph North
has few bright 24 \mum\ excess sources. The ``rest of the
cloud," largely concentrated to the northeast of L1688, shows
some intriguing small individual sources, but no YSO aggregates. 

Table 4 quantifies the degree of clustering seen by categorizing
the objects with 24 \mum\ excess. Consideration of all 24 \mum\ sources
with 2MASS associations shows that only 297 out of a total of 2590 sources are
confined to the L1688 subregion. This would imply a large distributed
population of potentially young sources in the cloud. However, when the 
selection criteria $K_s-[24]>2$,$K_s<14$ are applied, we see that
more than half of the YSO candidates are confined to L1688. About
60\% of the YSO candidates fall into the L1688, L1689, and L1709 
regions. The clustering of sources with Class I and flat SEDs is
more pronounced. A full 95\% of these sources are confined to
the central cluster and the L1709 and L1689 aggregates. In comparison,
only 61\% of the Class II candidates are located in these subregions.
It is worth noting that the nature of Class I and II candidates remain
unconfirmed by further study outside of the well-studied cluster 
and aggregates. Some percentage of these sources may well be extragalactic
interlopers.  Thus, star formation in the Ophiuchus clouds 
appears to be largely confined to 
the cluster and aggregates.  Small aggregates of the type seen in 
Taurus, Lupus, and Perseus are found in the L1709 and L1689 regions, but not 
in the remainder of the filamentary clouds in Ophiuchus.


\section{Conclusions}
\label{sec:concl}

We have presented the 14.4 square degree c2d {\it Spitzer} MIPS long
wavelength imaging survey of the Ophiuchus molecular cloud
complex. Many point sources and a variety of extended emission 
structures on scales
from circumstellar to intercloud are shown in our mosaicked
images. Using point-source fitting photometry, we have
identified and measured the brightness of 5779  sources at 24
\mum, 81 sources at 70 \mum, and 8 sources at 160 \mum. Using 
criteria derived from a K$_s$ versus K$_s$-[24] and other 
color-color and color-magnitude diagrams, we have
identified some 323 candidate young stellar objects throughout
the clouds. Several sources with extreme colors are discussed,
and a candidate edge-on disk is identified.  Our survey suggests
that although the numbers of forming stars are dominated by
the L1688 cluster and small aggregates in L1689 and L1709, 
there may be a population of Class II sources lurking
in the more tenuous parts of the cloud. This distributed population could
contain up to 40\% of the YSO candidates in Ophiuchus.


\acknowledgements 

Most of the support for this work, part of the Spitzer Space
Telescope Legacy Science Program, was provided by NASA through
contracts 1224608, 1230782, and 1230779, issued by the Jet
Propulsion Laboratory, California Institute of Technology under
NASA contract 1407.  We thank the Lorentz Center in Leiden for
hosting several meetings that contributed to this paper. 
Support for J.\ K.\ J.\ and P.\ C.\ M.\ was provided in part by
a NASA Origins grant, NAG5-13050.  Astrochemistry in Leiden is
supported by a NWO Spinoza grant and a NOVA grant.   K.\ E.\ Y.\
was supported by NASA under Grant No.\ NGT5-50401 issued through
the Office of Space Science. This research has made use of
NASA's Astrophysics Data System (ADS) Abstract Service, and of
the SIMBAD database, operated at CDS, Strasbourg, France.  This
research has made use of data products from the Two Micron
All-Sky Survey (2MASS), which is a joint project of the
University of Massachusetts and the Infrared Processing and
Analysis Center, funded by the National Aeronautics and Space
Administration and the National Science Foundation.  These data
were served by the NASA/IPAC Infrared Science Archive, which is
operated by the Jet Propulsion Laboratory, California Institute
of Technology, under contract with the National Aeronautics and
Space Administration.  The research described in this paper was
partially carried out at the Jet Propulsion Laboratory,
California Institute of Technology, under contract with the
National Aeronautics and Space Administration.

\clearpage

\begin{deluxetable}{llll}
\tablecaption{Summary of observations (Program 177)}
\tablewidth{0pt}
\tablehead{
\colhead{field} & \colhead{map center} & \colhead{first epoch
AORKEY} & \colhead{second epoch AORKEY} }
\startdata
ophi-mips1  & 16h45m50.0s,-21d21m16.0s & 5745408 & 5747200 \\
ophi-mips2  & 16h43m24.0s,-21d57m38.0s & 5745664 & 5747456 \\
1689-mips1  & 16h43m16.0s,-24d09m09.4s & 5748992 & 5753344 \\
1689-mips2  & 16h38m27.0s,-24d18m02.0s & 5749248 & 5753600 \\
1689-mips3  & 16h35m12.0s,-24d34m00.0s & 5749504 & 5753856 \\
1688-mips1  & 16h33m33.0s,-24d20m42.0s & 5757952 & 5766144 \\
1688-mips2  & 16h33m01.5s,-23d18m20.0s & 5758208 & 5766400 \\
1688-mips3  & 16h31m59.2s,-24d57m53.0s & 5758464 & 5766656 \\
1688-mips4  & 16h31m00.0s,-23d37m50.1s & 5758720 & 5766912 \\
1688-mips5  & 16h30m03.0s,-25d08m22.0s & 6605568 & 6605824 \\
1688-mips6  & 16h27m56.5s,-24d11m12.0s & 5759488 & 5767680 \\
1688-mips7  & 16h25m46.0s,-24d12m29.0s & 5759744 & 5676936 \\
1688-mips8  & 16h24m09.0s,-23d50m03.0s & 5760000 & 5768192 \\
1688-mips9  & 16h21m52.5s,-23d43m45.0s & 5760256 & 5768448 \\
oph-oc7     & 16h23m52.0s,-21d57m38.0s & 5778432 & 5778688 \\
oph-oc8     & 16h23m12.0s,-26d28m49.0s & 5778944 & 5779200 \\
\enddata
\end{deluxetable}

\clearpage

\begin{deluxetable}{ll}
\tablecaption{Box regions used to define individual clouds}
\tablewidth{0pt}
\tablehead{
\colhead{cluster} & \colhead{$\alpha$, $\delta$ of box corners} }
\startdata
L1688 &   16 30 12.0  -24 49 15.6, 16 25 05.5  -23 49 15.6\\
L1689 &   16 35 33.6  -25 30 00.0, 16 31 10.6  -24 10 37.2\\
L1709 &   16 35 33.6  -24 10 37.2, 16 30 12.0  -23 20 45.6\\
L1712 + L1729 &   16 46 28.1  -24 46 58.8, 16 35 33.6  -23 42 07.2\\
North &   16 48 10.3  -22 09 03.6,  16 42 21.6  -20 50 34.8\\
\enddata
\end{deluxetable}

\clearpage

\begin{deluxetable}{lrrrrrrrrr}
\rotate
\tabletypesize{\scriptsize}
\tablecaption{Detected Sources}
\tablewidth{0pt}
\tablehead{
\colhead{item} &  \colhead{overall} & \colhead{in IRAC cvg} &
\colhead{L1688} & \colhead{L1689}  & \colhead{L1709} &
\colhead{L1712+L1729} & \colhead{rest of cloud} &
\colhead{North} & \colhead{off-cloud}}
\startdata
number 24 \mum\ objects &             5779  & 2608 &569 & 490 &377&1082 & 2555 & 706 & 355 \\ 
number 70 \mum\ objects &              81   & 31& 24 &  12 &  5&12 & 14 & 13 & 2\\ 
number 160 \mum\ objects &              8   & 6&  1 &   1 &   2&1 & 1 & 2 & 0\\ 
number 24\&70 objects &                 50  & 31&  13 &  6 &   4&5 & 11& 11 & 0\\ 
number 24\&70\&160 objects &             5  & 4&  1 &   1 &   1&1 & 0 & 1 & 0\\ 
number 24 \mum\ \& K objects &         2590 & 1197&297 &  195 &  149&539 & 1102 & 308 & 177 \\ 
number 24 \mum\ \& any IRAC objects &  2306 & 2306&425 & 351 & 144&379&895& 112 & 0\\ 
number 70 \mum\ \& any IRAC objects &    42 & 30 & 23 &  7 &   4&3 &4&1 & 0\\ 
number 160 \mum\ \& any IRAC objects &   6  & 6&  1 &   1 &   2&1 &0&1 & 0\\ 
\enddata
\end{deluxetable}

\clearpage

\begin{deluxetable}{lllllllll}
\tablecaption{Classification based on \ks$-$[24]}
\rotate
\tabletypesize{\scriptsize}
\tablewidth{0pt}
\tablehead{
\colhead{item} & \colhead{entire cloud} &  \colhead{L1688} & 
\colhead{L1689}
& \colhead{L1709} & \colhead{L1712+L1729} & \colhead{rest of
cloud} & \colhead{North} & \colhead{off-cloud}}
\startdata
number objects with \ks\ and 24 \mum\ &2590      & 297 
& 195     & 149     &539      &1102     &308     &177 
\\
number with \ks$-$[24]$>$2, \ks$<$14  &323       & 161 
& 27      & 13      &18       &96       & 8      &2 
\\
number with \ks$-$[24]$>$2, \ks$<$14,
and Class I \ks$-$[24] color          &20 (6\%)  & 12 (7\%) 
& 2 (7\%) & 1  (7\%)& 0 (0\%) &5 (5\%)  & 0      &0 
\\
number with \ks$-$[24]$>$2, \ks$<$14,
and ``flat'' \ks$-$[24] color         &47 (14\%) & 33 
(20\%)& 3 (11\%)& 4 (30\%)& 1 (5\%) &5 (5\%)  &1 (12\%)&0 
\\
number with \ks$-$[24]$>$2, \ks$<$14,
and Class II \ks$-$[24] color         &207 (64\%)&100 
(62\%)&20 (74\%)& 7 (53\%)& 5 (27\%)&73 (76\%)&2 (25\%)&1 
(50\%)\\
number with \ks$-$[24]$>$2, \ks$<$14,
and Class III \ks$-$[24] color        &49 (15\%) & 16 (9\%) 
& 2 (7\%) & 1  (7\%)&12 (66\%)&13 (13\%)&5 (62\%)&1 (50\%)\\
\enddata
\end{deluxetable}

\clearpage

\begin{deluxetable}{lll}
\tablecaption{IRAS Results in MIPS Ophiuchus Map (main+North)}
\tablewidth{0pt}
\tablehead{
\colhead{item} & \colhead{\ldots\ at 24  } & \colhead{\ldots\
at 70 \mum} }
\startdata
12 \mum\ PSC real (IRAS qual=3) detections & 100  &  91           \\
\hskip 36pt cleanly retrieved  &            74 (74\%)    &  25 (27\%)     \\
\hskip 36pt completely missing  &            2 ( 2\%)    &  41 (45\%)    \\
\hskip 36pt nebulosity; no point source  &15 (15\%)    &  24 (26\%)\\
\hskip 36pt resolved as multiple  &          9 (9\%)    &   1 ( 1\%)    \\
25 \mum\ PSC real (IRAS qual=3) detections &63  &  63            \\
\hskip 36pt cleanly retrieved  &           41 (65\%)     &   26 (41\%)     \\
\hskip 36pt completely missing  &            1 ( 1\%)    &    12 (19\%)    \\
\hskip 36pt nebulosity; no point source &           15 (23\%)&        24 (38\%)\\
\hskip 36pt resolved as multiple  &            6 (9\%)    &     1 ( 1\%)    \\
60 \mum\ PSC real (IRAS qual=3) detections &       93  &  84           \\
\hskip 36pt cleanly retrieved  &           23 (24\%)     &   20 (23\%)     \\
\hskip 36pt completely missing  &           41 (44\%)    &    16 (19\%)    \\
\hskip 36pt nebulosity; no point source  &           27 (29\%)&  48 (57\%)\\
\hskip 36pt resolved as multiple  &            2 (2\%)    &     0 ( 0\%)    \\
100 \mum\ PSC real (IRAS qual=3) detections &       55 &   50          \\
\hskip 36pt cleanly retrieved  &           4 (7\%)     &    3 (6\%)     \\
\hskip 36pt completely missing  &           29 (52\%)    &    12 (24\%)    \\
\hskip 36pt nebulosity; no point source  &           22 (40\%)&    35 (70\%)\\
\hskip 36pt resolved as multiple &            0 (0\%)    &     0 ( 0\%)    \\
\hline
12 \mum\ FSC real (IRAS qual=3) detections &        50  &  35           \\
\hskip 36pt cleanly retrieved  &           42 (84\%)     &   3 (8\%)     \\
\hskip 36pt completely missing  &            0 (0\%)    &    24 (68\%)    \\
\hskip 36pt nebulosity; no point source  &            7 (14\%)&  8 (22\%)\\
\hskip 36pt resolved as multiple  &            1 (2\%)    &     0 (0\%)    \\
25 \mum\ FSC real (IRAS qual=3) detections &           14  &  11            \\
\hskip 36pt cleanly retrieved  &           13 (92\%)     &   5 (45\%)     \\
\hskip 36pt completely missing  &            0 ( 0\%)    &    6 (54\%)    \\
\hskip 36pt nebulosity; no point source  &            0 (0\%)&        0 (0\%)\\
\hskip 36pt resolved as multiple &            1 (7\%)    &     0 (0\%)    \\
60 \mum\ FSC real (IRAS qual=3) detections &            0  &   0            \\
100 \mum\ FSC real (IRAS qual=3) detections &            0& 0\\
\enddata
\end{deluxetable}

\clearpage

\begin{deluxetable}{lll}
\tablecaption{IRAS PSC detections not recovered by Spitzer/MIPS}
\tablewidth{0pt}
\tablehead{
\colhead{PSC name} &  \colhead{\ldots from 24 \mum} &
\colhead{\ldots from 70 \mum}}
\startdata
16169-2443&missing (upper limit at 12,25)&missing (det@60, upp lim@100)\\ 
16172-2503&missing (upper limit at 12,15)&confused by nebulosity\\ 
16173-2422&confused by nebulosity&missing (det@60, upp lim@100)\\ 
16184-2452&confused by nebulosity&confused by nebulosity\\ 
16184-2452&confused by nebulosity&confused by nebulosity\\ 
16187-2339&confused by nebulosity&missing (upper limit at 60,100)\\ 
16193-2335&confused by nebulosity&confused by nebulosity\\ 
16193-2401&confused by nebulosity&confused by nebulosity\\ 
16193-2450&confused by nebulosity&confused by nebulosity\\ 
16194-2410&missing (upper limit at 12,25)&confused by nebulosity\\ 
16194-2500&off edge&confused by nebulosity\\ 
16196-2443&missing (upper limit at 12,25)&confused by nebulosity\\ 
16200-2251&confused by nebulosity&confused by nebulosity\\ 
16201-2410&retrieved (upper limit at 12)&confused by nebulosity\\ 
16202-2356&missing (upper limit at 12,25)&confused by nebulosity\\ 
16202-2427&missing (upper limit at 12,25)&missing (det@60, upp lim@100)\\ 
16203-2332&missing (upper limit at 12,25)&missing (det@60, upp lim@100)\\ 
16205-2308&missing (upper limit at 12,25)&confused by nebulosity\\ 
16209-2422&confused by nebulosity&confused by nebulosity\\ 
16212-2316&missing (upper limit at 12,25)&missing (det@60, upp lim@100)\\ 
16214-2302&missing (upper limit at 12,25)&confused by nebulosity\\ 
16214-2345&confused by nebulosity&confused by nebulosity\\ 
16214-2436&confused by nebulosity&confused by nebulosity\\ 
16217-2244&missing (upper limit at 12,25)&confused by nebulosity\\ 
16219-2344&confused by nebulosity&confused by nebulosity\\ 
16219-2417&confused by nebulosity&confused by nebulosity\\
16221-2428&confused by nebulosity&confused by nebulosity\\ 
16222-2358&confused by nebulosity&confused by nebulosity\\ 
16223-2404&confused by nebulosity&confused by nebulosity\\ 
16225-2417&confused by nebulosity&confused by nebulosity\\ 
16226-2319&broken into pieces&confused by nebulosity\\ 
16226-2420&confused by nebulosity&confused by nebulosity\\ 
16227-2418&confused by nebulosity&confused by nebulosity\\ 
16228-2432&confused by nebulosity&confused by nebulosity\\ 
16229-2413&confused by nebulosity&confused by nebulosity\\
16233-2421&confused by nebulosity&confused by nebulosity\\ 
16234-2436&confused by nebulosity&confused by nebulosity\\ 
16234-2436&confused by nebulosity&confused by nebulosity\\ 
16235-2416&retrieved&confused by nebulosity\\ 
16238-2317&missing (upper limit at 12,25)&confused by nebulosity\\ 
16246-2440&retrieved&confused by nebulosity\\ 
16257-2533&missing (upper limit at 12,25)&confused by nebulosity\\ 
16262-2545&retrieved&missing (det@60, upp lim@100)\\ 
16265-2350&missing (upper limit at 12,25)&confused by nebulosity\\ 
16267-2539&confused by nebulosity&confused by nebulosity\\ 
16268-2533&missing (upper limit at 12,25)&missing (det@60, upp lim@100)\\ 
16269-2454&confused by nebulosity&confused by nebulosity\\ 
16272-2541&confused by nebulosity&confused by nebulosity\\ 
16274-2349&confused by nebulosity&confused by nebulosity\\ 
16275-2251&missing (upper limit at 12,25)&missing (det@60, upp lim@100)\\ 
16277-2356&confused by nebulosity&missing (det@60, upp lim@100)\\ 
16280-2353&confused by nebulosity&missing (det@60, upp lim@100)\\ 
16281-2514&missing (upper limit at 12,25)&confused by nebulosity\\ 
16285-2519&missing (upper limit at 12,25)&confused by nebulosity\\ 
16289-2527&missing (upper limit at 12,25)&confused by nebulosity\\ 
16301-2525&missing (upper limit at 12,25)&confused by nebulosity\\ 
16303-2428&confused by nebulosity&confused by nebulosity\\ 
16304-2504&missing (upper limit at 12,25)&confused by nebulosity\\ 
16306-2435&confused by nebulosity&confused by nebulosity\\ 
16311-2419&confused by nebulosity&confused by nebulosity\\ 
16313-2439&confused by nebulosity&confused by nebulosity\\ 
16318-2519&missing (upper limit at 12,25)&missing (det@60, upp lim@100)\\ 
16322-2421&confused by nebulosity&confused by nebulosity\\ 
16325-2433&confused by nebulosity&confused by nebulosity\\ 
16330-2424&missing (upper limit at 12,25)&confused by nebulosity\\ 
16330-2431&missing (upper limit at 12,25)&confused by nebulosity\\ 
16335-2402&missing (upper limit at 12,25)&missing (det@60, upp lim@100)\\ 
16335-2419&missing (upper limit at 12,25)&confused by nebulosity\\ 
16339-2422&missing (upper limit at 12,25)&confused by nebulosity\\ 
16341-2413&missing (upper limit at 12,25)&missing (det@60, upp lim@100)\\ 
16342-2413&missing (upper limit at 12,25)&missing (det@60, upp lim@100)\\ 
16375-2439&missing (upper limit at 12,25)&confused by nebulosity\\ 
16379-2444&missing (upper limit at 12,25)&missing (det@60, upp lim@100)\\ 
16383-2407&missing (upp lim@12, det@25)&missing (upper limit at 60,100)\\ 
16414-2337&missing (upper limit at 12,25)&missing (det@60, upp lim@100)\\ 
16417-2155&confused by nebulosity&confused by nebulosity\\ 
16418-2155&confused by nebulosity&confused by nebulosity\\ 
16421-2146&confused by nebulosity&confused by nebulosity\\ 
16426-2129&confused by nebulosity&confused by nebulosity\\ 
16436-2121&confused by nebulosity&confused by nebulosity\\ 
\enddata
\end{deluxetable}

\clearpage

\begin{deluxetable}{llllllll}
\rotate
\tabletypesize{\scriptsize}
\tablecaption{MIPS photometry for point-source objects discussed in this paper}
\label{tab:objects}
\tablewidth{0pt}
\tablehead{
\colhead{SSTc2d name} & \colhead{region} &
\colhead{MIPS-24 (mJy)\tablenotemark{a}} & \colhead{MIPS-70 (mJy)\tablenotemark{a}} &
\colhead{MIPS-160 (mJy)\tablenotemark{a}} & \colhead{other
names\tablenotemark{b}} &
\colhead{notes} }
\startdata
\cutinhead{Red sources selected by a variety of means (see text)}
162145.1-234231&rest of cloud&  206 &  730 &   \ldots & [L89] R7 &  \\
162218.5-232148&rest of cloud&  804 &  770 &   \ldots & IRAS 16193-2314=WSB 12& \\
162309.2-241704&rest of cloud&  715 & 1400 &   \ldots & IRAS 16201-2410 & \\
162502.0-245932&rest of cloud&  395 &  630 &   \ldots & IRAS 16220-2452=WSB 19 & \\
162510.5-231914&rest of cloud&  840 &  860 &   \ldots & IRAS 16221-2312=V2503 Oph=DoAr 16=WSB 20=HBC 257& \\
162524.3-242756&  L1688 & 127 &   \ldots &   \ldots & HD 147889 & discussed in \S4.1.1 \\
162623.5-242439&  L1688 & 927 &   \ldots &   \ldots & CRBR 2322.1-1754 (=possibly ISO-Oph 37)& discussed in \S4.1.1 \\
162646.4-241200&  L1688&  287 &  250 &   \ldots & YLW 37& \\
162648.4-242838&  L1688 & 455 &   \ldots &   \ldots & CRBR 2346.8-2157 (=possibly ISO-Oph 70=WL 2) & discussed in \S4.1.1 \\
162706.7-243814&  L1688& 2710 & 3600 &   \ldots & BBRCG 22=ISO-Oph 103=WL 17 & \\
162713.8-244331&  L1688&  318 &  550 &   \ldots & ISO-Oph 117=WLY 2-32b & \\
162717.5-242856&  L1688 & 855 &   \ldots &   \ldots & ISO-Oph 124=YLW 12A & discussed in \S4.1.1 \\
162737.2-244238&  L1688&  576 &  890 &   \ldots & (=possibly ISO-Oph 161 - 6\arcsec\ away) & discussed in \S4.1.1 \\
162738.3-243658&  L1688&  702 &  930 &   \ldots & IRAS 16246-2430=ISO-Oph 163=WLY 1-47=WLY 2-49& \\
162738.9-244020&  L1688&  474 & 1200 &   \ldots & ISO-Oph 165 & \\
162739.4-243915&  L1688& 1160 & 1400 &   \ldots & ISO-Oph 166=WSB 52& \\
162739.8-244315&  L1688& 2520 & 2900 &   \ldots & IRAS 16246-2436=ISO-Oph 167=WLY 2-51=WLY 1-45& \\
162741.6-244644&  L1688&   72 &  320 &   \ldots & ISO-Oph 170& \\
162816.5-243657&  L1688&  239 &  620 &   \ldots & (=possibly ISO-Oph 196=WSB 60 - 8\arcsec\ away) & \\
162821.6-243623&  L1688&   94 & 2100 & 9910 & [SSG2006] MMS126&   \\
162845.2-242819&  L1688&  737 & 1100 &   \ldots & V853 Oph=DoAr 40=IRAS 16257-2421=ISO-Oph 199=WSB 62 & \\
162857.8-244054&  L1688&  167 &  240 &   \ldots & (=possibly BKLT J162858-244054 - 7\arcsec\ away) & \\
163130.8-242439&  L1689& 1250 &  900 & 4690 & IRAS 16284-2418=DoAr 43=WSB 71 & \\
163133.4-242737&  L1689& 1430 & 2600 &   \ldots & IRAS 16285-2421=DoAr 44=SVS 2114=HBC 268=WSB 72 &  \\
163133.8-240446&  L1709&  717 &  440 &   \ldots & IRAS 16285-2358 & \\                              
163135.6-240129&  L1709& 2980 &10647 &19000 & IRAS 16285-2355=WLY 2-63 & \\
163136.7-240420&  L1709&  194 &  670 &   \ldots & \ldots& \\
163143.7-245524&  L1689& 1210 & 1400 &   \ldots & ISO-Oph 200& \\
163144.5-240212&  L1709&  275 &  390 &   \ldots & \ldots& \\
163152.4-245536&  L1689&  887 &  680 &   \ldots & ISO-Oph 203 & \\
163154.7-250323&  L1689& 1380 & 1400 &   \ldots & WSB 74 (=possibly IRAS 16289-2457)& \\
163222.6-242832&  L1689 & 595 &   \ldots &   \ldots & IRAS 16293-2422 & discussed in \S4.1.1 \\
163355.6-244204&  L1689&  235 &  740 &   \ldots & RX J1633.9-2442=[NWA2006] SMM17  & \\
163528.6-245648&  L1689 & \ldots &  740.0 &   \ldots & CD-24 12715 & debris disk candidate; see \S4.1.1\\
163945.4-240203& L1712+L1729& 1420 & 2000 & 4560 & IRAS 16367-2356=WSB 82& \\
164417.8-220648&  north& 1240 &  220 &   \ldots & IRAS 16413-2201 & debris disk candidate; see \S4.1.1 \\
164526.1-250316&rest of cloud&  150 &  260 &   \ldots & IRAS 16424-2457 & \\
\cutinhead{L1709 aggregate}
163129.2-240431&  L1709&  82.2 &   \ldots &   \ldots && src 1 in \S4.1.2\\
163133.8-240446&  L1709&  717.0 &  440.0 &   \ldots  & IRAS 16285-2358 & src 2 in \S4.1.2\\
163134.2-240325&  L1709&  747.0 &   \ldots &18300.0  && src 3 in \S4.1.2\\
163135.6-240129&  L1709& 2980.0 &10647.0 &19000.0 & IRAS 16285-2355 & src 4 in \S4.1.2\\     
163136.7-240420&  L1709&  194.0 &  670.0 &   \ldots && src 5 in \S4.1.2, new edge-on disk candidate\\
163144.5-240212&  L1709&  275.0 &  390.0 &   \ldots && src 6 in \S4.1.2 \\
\cutinhead{L1689 aggregate}
163143.7-245559& L1689&   19.9 &   \ldots &   \ldots & & src 1 in \S4.1.3\\
163143.7-245524& L1689& 1210.0 & 1400.0 &   \ldots & ISO-Oph 200&src 2 in \S4.1.3 \\
163148.9-245429& L1689&   20.9 &   \ldots &   \ldots & ISO-Oph 201&src 3 in \S4.1.3 \\
163151.9-245623& L1689& \ldots & 1500.0 &  \ldots & (=possibly ISO-oph 204=LDN 1689 IRS 5) & src 4 in \S4.1.3; saturated in nearly all bands\\
163152.0-245725& L1689&   39.8 &   \ldots &   \ldots & & src 5 in \S4.1.3; unresolved binary \\
163152.4-245536& L1689&  887.0 &  680.0 &   \ldots & ISO-Oph 203&src 6 in \S4.1.3 \\
163153.4-245505& L1689&  104.0 &   \ldots &   \ldots & ISO-Oph 205&src 7 in \S4.1.3 \\
163159.3-245440& L1689&   43.8 &   \ldots &   \ldots & ISO-Oph 208&src 8 in \S4.1.3 \\
163200.9-245641& L1689&   \ldots &46800.0 &   \ldots & IRAS 16289-2450=LDN 1689S& src 9 in \S4.1.3 \\
\enddata
\tablenotetext{a}{Absolute uncertainties on the 24 \mum\ data are
estimated to be 10-15\%; statistical uncertainties are much less
than this.  Uncertainties on 70 and 160 \mum\ flux densities are
estimated to be 20\%.}
\tablenotetext{b}{Source associations taken from SIMBAD up to
5\arcsec\ away from Spitzer position (known to a fraction of an
arcsecond); possible associations listed in SIMBAD as being
5-8\arcsec\ away are listed as such.  In some cases, there are more
than a dozen names for these objects, so only the most common are
listed.   References for these catalogued sources include Loren (1989;
L89), Wilking et al.\ (2005; WSB), Barsony et al.\ (1989; BBRCG),
Wilking et al.\ (1989; WLY), Stanke et al.\ (2006; SSG2006),  Barsony
et al.\ (1997; BKLT), Nutter et al.\ (2006; NWA2006), Dolidze and
Arakelian (1959; DoAr), and Bontemps et al.\ (2001; ISO-Oph).}
\end{deluxetable}

\clearpage

\clearpage

\clearpage

\clearpage

\clearpage

\clearpage

\clearpage

\clearpage

\clearpage

\clearpage

\clearpage

\clearpage

\clearpage

\clearpage

\clearpage

\clearpage

\clearpage

\clearpage

\clearpage

\clearpage

\clearpage

\clearpage

\end{document}